\documentstyle[pre,aps,multicol,psfig]{revtex}
\def\openone{\leavevmode\hbox{\small1\kern-3.3pt\normalsize1}}
\begin{document}
\draft

\title{The Impact of Isospin Breaking on the Distribution 
           of Transition Probabilities}

\author{C.I. Barbosa, T. Guhr and H.L. Harney}

\address{Max--Planck--Institut f\"ur Kernphysik, 
         Postfach 103980,
         69029 Heidelberg, 
         Germany}

\date{\today}

\maketitle              

\begin{abstract}
  In the present paper we investigate the effect of symmetry breaking
  in the statistical distributions of reduced transition amplitudes
  and reduced transition probabilities.  These quantities are easier
  to access experimentally than the components of the eigenvectors and
  were measured by Adams et al.~for the electromagnetic transitions in 
  $^{26}$Al.  We focus on isospin symmetry breaking described by 
  a matrix model where both, the
  Hamiltonian and the electromagnetic operator, break the symmetry.
  The results show that for partial isospin conservation, the
  statistical distribution of the reduced transition probability
  can considerably deviate from the Porter--Thomas distribution.
\end{abstract} 


\begin{multicols}{2}


\narrowtext  

\section{Introduction}
\label{I}

The spectral fluctuations in a rich variety of different physical
systems show, if measured on the scale of the local mean level
spacing, very similar features. This high degree of universality makes
it possible to describe these fluctuations with random matrices.
Random Matrix Theory (RMT) is a simple, schematic model in which the
matrix elements of the Hamiltonian in some basis are replaced
with random numbers. Apart from randomness, the only further input are
the symmetries and invariances of the system, in particular
time reversal invariance. It turns out that this assumption of full
ergodicity or ``chaoticity'' leads, in many cases, to a complete and
parameter free description of the spectral fluctuations. Such
universal cases are said to be of Wigner--Dyson type. We refer the
reader to the reviews in Refs.~\cite{Wigner,1}.  Originally, Wigner
had developed this approach in nuclear physics where it continues to
find new applications. 

In recent years, interest has been focussed on deviations
from the universal, parameter free Wigner--Dyson fluctuations. The deviations
can have different reasons, such as regular effects competing with
full chaoticity or the breaking of time reversal invariance, see
Ref.~\cite{1}. Here, we wish to discuss deviations attributed to the breaking
of isospin symmetry. We recall that symmetries such as isospin or
parity are, in contrast to time reversal invariance, associated with
quantum numbers. In the context of spectral fluctuations and RMT,
symmetry breaking already received broad interest, see the compilation
in Ref.~\cite{1}. As early as 1960, Rosenzweig and
Porter~\cite{Porter} analyzed atomic spectra by setting up a general
random matrix model which describes crossover transitions between
different angular momentum coupling schemes. In the late eighties,
Mitchell et al.~\cite{values} measured and analyzed about 100 low
lying states with known values of the isospin quantum number in the
nucleus $^{26}$Al. In Ref.~\cite{Weid}, this was discussed in the
framework of a random matrix model which is a special case of the
Rosenzweig--Porter model. An estimate for the statistical Coulomb
matrix element, i.e.~a measure for the degree of isospin breaking,
could be obtained.  Motivated by similar questions in molecular
physics, Leitner and coworkers~\cite{Leitner} performed a perturbative
calculation of the spectral fluctuations in the random matrix model.
More recently, new data were obtained in two statistically highly
significant experiments on the breaking of a point group symmetry in a
resonating quartz block~\cite{cristais2}, and the coupling of two
chaotic microwave billiards \cite{nos}. Both cases, although
physically very different, are statistically fully equivalent to
symmetry breaking in quantum mechanics.  Importantly, there is only
one parameter entering the random matrix model. It is a unique
measure e.g.~for the root mean square statistical Coulomb matrix element.
This illustrates that the random matrix model is the
ideal tool to extract a root mean square symmetry breaking matrix
element from the data.

All these studies addressed the {\it spectral} fluctuation properties.
Symmetry breaking, however, will also have an impact on the statistics
of the {\it wave functions} and observables sensitive to them.
Recently, Adams, Mitchell and Shriner~\cite{26Al} collected reduced
$\gamma$ ray transition probabilities from different experiments on
$^{26}$Al. As mentioned above, this nucleus had already shown a strong
deviation of the {\it spectral} fluctuations from the universal
Wigner--Dyson result due to isospin breaking~\cite{values}.  The new
results~\cite{26Al} show that the distribution of the {\it transition}
probabilities also considerably deviates from the Wigner--Dyson
statistics, i.e.~from the distribution that corresponds to full,
parameter free ``chaos''.  In the present contribution, we wish to
discuss these results. To this end, we extend the random matrix model
of Ref.~\cite{Weid} to discuss transition probabilities. Similar
investigations were performed simultaneously and independently by Andersen,
Ellegaard, Jackson and Schaadt~\cite{KBH} for symmetry breaking in acoustic
and elastomechanical systems.

The article is organized as follows. In Sec.~\ref{II} we briefly
review the experimental results on the reduced transition
probabilities in $^{26}$Al collected by Adams, Mitchell and
Shriner~\cite{26Al}.  In Sec.~\ref{III}, we discuss the random matrix
model in the case that no symmetry is present, in particular its
predictions for transition probabilities. The random matrix model for
isospin breaking is numerically studied in Sec.~\ref{IV}. In Sec.~\ref{V}, the
numerical simulation is approximated by a qualitative albeit analytical model. 
The data analysis
is performed in Sec.~\ref{VI}. Summary and conclusion are presented in
Sec.~\ref{VII}.

\section{Experimental results}
\label{II}

Experimental reduced electromagnetic transition strengths between 
the excited states of the nucleus $^{26}$Al have been collected by 
Adams et al. --- see Ref.~\cite{26Al}. 
Their data involve levels between the ground
state and the excitation energy of 8.067 MeV. In this region, states
with isospin $T=0$ and $T=1$ are found and isospin is known to be
approximately conserved.

The probability $B_{if}$ of a transition from the initial
configuration $|i\rangle $ to the final configuration 
$|f\rangle $
is the square 
\begin{eqnarray}
B_{if} = |W_{if}|^2
\label{eqII.3}
\end{eqnarray}
of the matrix element
\begin{eqnarray}
W_{if} = \langle f|{\cal{O}}|i\rangle
\label{eqII.4}
\end{eqnarray}
of the relevant transition operator ${\cal{O}}$ in a special basis.

Approximately 180 levels and 1500 electromagnetic transitions are
known. The fluctuations of the $B_{if}$ values shall be studied. To this
end, their systematic dependence on the quantum numbers of the initial
and final states must be removed. This has been done in Ref.~\cite{26Al} in
the following way. The states in $^{26}$Al have been characterized by
their excitation energy $E$, spin $J$, parity $\pi $, and isospin $T$.
Transitions are characterized by their electromagnetic character $X$ 
which may be 
${\cal E}$ or ${\cal M}$, their multipolarity $L$ and a label $\tau $ which
becomes isoscalar $(IS)$ if $\Delta T=0$ and isovectorial $(IV)$ if
$\Delta T =1$. Hence, both the transition operator ${\cal O} = {\cal O} (XL,
\tau )$ and the transition probabilities $B_{if} = B_{if} (XL, \tau )$ 
are functions of $X$, $L$ and $\tau $. We shall, however, not always
write all these arguments.

A {\it transition sequence} is defined as a set of reduced
transition probabilities where the initial states 
have a common assignment $J^{\pi }$, $T$,
the final states have a common assignment $J^{\prime \pi ^{\prime}}$, 
$T^{\prime }$, and the transitions have their three defining
characteristics all the same. Thus the reduced transition
probabilities $B_{if} = B(E_{i},E_{f})$ of a given transition sequence can
be labeled by the energies $E_{i}$ and $E_{f}$ of the initial and
final states, respectively. The aforementioned secular variations of
the $B$ values were removed by normalizing them to the local average
value $\langle B(E_{i},E_{f}) \rangle $ of $B(E_{i},E_{f})$ so that
the statistical variable used further on is
\begin{eqnarray}
  y(E_{i},E_{f}) = \frac{B(E_{i},E_{f})}
  {\langle B(E_{i},E_{f}) \rangle } .
\label{eqII.1}
\end{eqnarray}
The local average is defined by help of weighting factors that are
Gaussian functions of the excitation energies. In doing so one must,
however, remove the systematic dependence of the local level distance
$D$ on the excitation energy; i.e.~one measures the energy in units of
$D$ and works with a dimensionless energy
\begin{eqnarray}
  \varepsilon = \frac{E}{D} .
\label{eqII.2}
\end{eqnarray}
The details are given in Ref.~\cite{22Na}. The local average of the $B$
values of a given transition sequence is then
\begin{eqnarray}
  {\langle B (\varepsilon _{i} , \varepsilon _{f}) \rangle} = 
  \frac 
  {\sum _{\varepsilon ,\varepsilon ^{\prime }} 
  B (\varepsilon , \varepsilon ^{\prime }) \,\,
  e^{-(\varepsilon _{i}-\varepsilon )^{2} /8} 
  e^{-(\varepsilon _{f}- \varepsilon ^{\prime })^{2}/8} } 
  {\sum _{\varepsilon ,\varepsilon ^{\prime }}
  e^{-(\varepsilon _{i}-\varepsilon )^{2} /8} 
  e^{-(\varepsilon _{f}-\varepsilon ^{\prime })^{2}/8} } .
\label{eqIVA.1}
\end{eqnarray}
The fact that the variance of the Gaussian functions has been chosen
equal to 4 is discussed in Ref.~\cite{22Na}.

It is clear that the definition of the quantities $y=y(E_{i},E_{f})$ 
requires the
necessary spectroscopic information for the relevant states in
$^{26}$Al. Furthermore, the local averages $D$ and $\langle B(E_{i},
E_{f} )\rangle $ require that there is a minimum number of
members in the sequences from which these averages are derived. These
restrictions have finally led in Ref.~\cite{26Al} to an ensemble of 873
values of $y$.

\begin{figure} [hbt]
\centerline{\psfig{figure=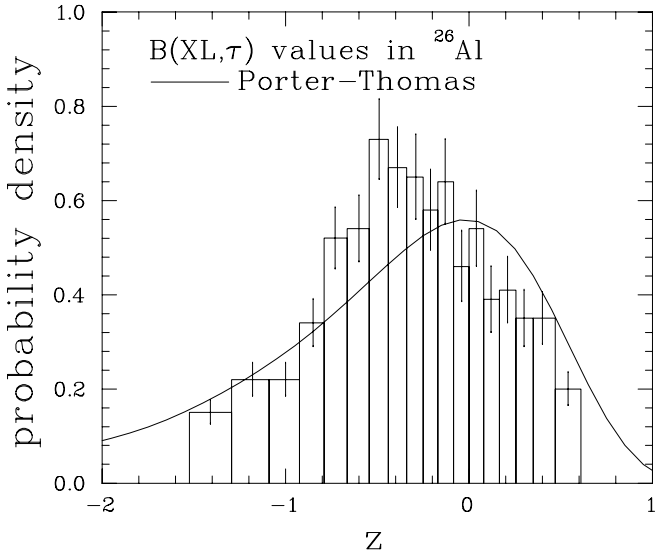,%
width=8.6cm,bblly=3.5cm,bburx=19.4cm,bbllx=1cm,bbury=24.4cm,angle=90}}
\caption{\protect The distribution of experimental reduced transition 
         probabilities $B(XL,\tau )$ in $^{26}$Al from Ref.[9]. The quantity $B$
	 has been transformed to the logarithmic variable $z$ of Eq.~(6). 
	 The curve labeled
	 Porter-Thomas is the distribution~(9).}
\label{fig26Al}
\end{figure}

The authors of Ref.~\cite{26Al,22Na} have found it convenient to transform
$y$ to the logarithmic variable
\begin{eqnarray}
z = \log _{10} y \ .
\label{eqII.2a}
\end{eqnarray}
The experimental distribution of $z$ is given by the histogram on
Fig.~\ref{fig26Al}. Although the widths $\Delta z_{k}$ of the bins 
$k=1 \cdots 18$
in Fig.~\ref{fig26Al} vary, the ordinate is the probability density with respect
to $z$: The height $p_{k}$ of the $k$-th bin is
\begin{eqnarray}
p_{k} \sim \frac{N_{k}}{\Delta z_{k}} \ ,
\label{eq6a}
\end{eqnarray}
where $N_{k}$ is the number of cases falling  into the $k-$th bin.

Since the states
$|i\rangle$ and $|f\rangle$ are believed to be ``very complicated", it
is natural to assume that the matrix elements~(\ref{eqII.4}) have a
Gaussian distribution; see the discussion in Sec.~\ref{IIIB}. 
This entails that $y$ has a Porter--Thomas
distribution with unit average value, i.e.
\begin{eqnarray}
  P(y) = \frac{1}{\sqrt{2\pi}} \frac{\exp(-y/2 )}{\sqrt{y}} .
\label{eqII.5} 
\end{eqnarray} 
Transforming this to the variable $z$ yields
\begin{eqnarray}
  P(z) = \frac{\ln\,10}{\sqrt{2\pi}}\,{10^{z/2}}\,\exp(-10^z/2) .
\label{eqII.6}
\end{eqnarray}
A comment on the notation is in order: For every probability density we
shall use the symbol $P$. The fact that we deal with different
functions will be clearly indicated by the argument of the probability
density in question.

The distribution~(\ref{eqII.6}) is given on Fig.~\ref{fig26Al} 
for comparison with
the data. The distribution~(\ref{eqII.6}) is normalized to unity; the histogram of
the data, however, has been normalized to 
\begin{eqnarray}
\sum _{k=1}^{18} p_{k} \Delta z_{k} =0.83
\label{8a}
\end{eqnarray}
by the authors of Ref.~\cite{26Al}. This is motivated \cite{JS} by the fact that
$\int dz \, P(z)$ over the range of the data amounts to 0.83 and
approximately takes care of the upper and lower detection thresholds.

The point of Ref.~\cite{26Al} is the shift of the experimental
distribution with respect to the Porter--Thomas distribution
$P (z)$. The maximum of the histogram occurs around $z=-0.5$. The
maximum of $P(z)$ occurs at $z=0$. The authors of Ref.~\cite{26Al} 
have conjectured
that the discrepancy between the experimental data and the
Porter--Thomas distribution may be a consequence of isospin breaking
in $^{26}$Al. The following sections of the present paper focus on
this conjecture. In the next section, a random matrix model of isospin
violation --- and more generally of symmetry breaking --- is
presented.

\section{Random Matrix Model --- the case of no symmetry}
\label{III}

In Sec.~\ref{IIIA}, we briefly review the distributions
of wave functions and widths for the standard Wigner--Dyson case in
which no symmetry is present. A more detailed presentation with
further references can be found in Ref.~\cite{1}.  In Sec.~\ref{IIIB},
we work out the distribution of the transition matrix elements for the
same case.

\subsection{Wave functions and decay amplitudes}
\label{IIIA}

If the system is invariant under time reversal, the wave functions
can be chosen real and the $N \times N$ random matrix $H$ modeling the
Hamiltonian is real and symmetric. The matrix elements are Gaussian
distributed random numbers and $H$ is said to be in the Gaussian
Orthogonal Ensemble (GOE). In the physically relevant limit
of infinitely many states, $N\to\infty$, the fluctuation properties
are of Wigner--Dyson type. A wave function is modeled by an
eigenvector $u_i, \ i=1 \cdots N$ of $H$, i.e.~we have $Hu_i=E_i u_i$
where $E_i$ is the eigenenergy.

We are interested in the probability density $P(a)$ of finding
the value $a$ with $-1\le a\le +1$ for a component $u_{im}$ of the 
eigenvector $u_i$. For finite $N$ one finds
\begin{eqnarray} 
  P_N(a) = \frac{\Gamma(N/2)}{\sqrt{\pi }\Gamma ((N-1)/2)} (1-
  a^{2}) ^{(N-3)/2} \ .
\label{eqIII.0}
\end{eqnarray}
The second moment of this distribution is 
$\overline{a^2}=1/N$. For a large number of levels one obtains a
Gaussian with variance $1/N$, i.e.
\begin{eqnarray}
  P_N(a)&=&\sqrt{\frac{N}{2\pi}} \exp\left(-\frac{N}{2}a^2\right) 
            \nonumber \\
        &=& G(a,N^{-1/2}) \ .
\label{eqIII.1}
\end{eqnarray} 
Here, we have introduced the notation $G(a,\sigma )$ for a Gaussian with
variance $\sigma ^{2}$.

The wave functions are rarely accessible in an experiment. However,
other observables, such as scattering matrix elements or partial
widths, sensitively depend on them. Consider the scattering from a
state $i$ with wave function $u_i$ into a channel $c$ with channel
wave function $\chi_c$.  The corresponding reduced partial width
amplitude $\gamma_{ic}$ can be written as
\begin{eqnarray}
\gamma_{ic} = \sum_{m=1}^{N} u_{im}J_{mc}  ,
\label{extraIII.1}
\end{eqnarray}
where $J_{mc}$ is the overlap integral between the $m-$th canonical 
basis vector
and $\chi_c$. Again, the probability density can be worked out
\begin{eqnarray}
  P(\gamma) &=& \frac{\Gamma(N/2)} {\left(\pi
    N\overline{\gamma^2}\right)^{1/2}\Gamma((N-1)/2)}
                   \nonumber\\
   & & \qquad\qquad\qquad \times
  \left(1-\frac{\gamma^2}{N\overline{\gamma^2}}\right)^{(N-3)/2} \ ,
\label{eqIII.2}
\end{eqnarray} 
where we suppress the indices $i$ and $c$. The second moment reads
$\overline{\gamma^2}=\overline{\gamma_{ic}^2}=N^{-1}\sum_{m=1}^NJ_{mc}^2$.
We notice that the functional form of this distribution agrees with
that of Eq.~(\ref{eqIII.0}). Thus, in the limit of large N,
one again finds a Gaussian. Usually, one introduces the partial width
$\Gamma = \gamma^2$ which can be measured. The relative partial width
$y=\Gamma/\overline{\Gamma}$ with
$\overline{\Gamma}=\overline{\gamma^2}$ is distributed according to
the Porter--Thomas law~(\ref{eqII.5}).

We notice that the Porter--Thomas law or, equivalently, the Gaussian
for the partial width amplitudes $\gamma$, results from a large $N$
expansion of the distribution~(\ref{eqIII.2}). 
Alternatively, one may derive these large $N$ results by
using the central limit theorem: The partial width amplitudes are,
according to Eq.~(\ref{extraIII.1}), given as a linear combination of
$N$ components $u_{im}$. If --- as assumed in~(\ref{eqIII.2}) --- 
all these components are independently
distributed, the distribution of the partial width amplitudes
approaches a Gaussian for large $N$. This line of arguing does not use
the fact that the distribution of every single component $u_{im}$ is
Gaussian for large $N$. It would apply for any smooth distribution of
$u_{im}$,
provided it does not sensitively depend on $N$. This subtle point will
be important in Sec.~\ref{IVA}.

\subsection{Transition matrix elements}
\label{IIIB}

The results compiled in Sec.~\ref{IIIA} apply to the partial widths and
to their amplitudes. However, in the experiments on $^{26}$Al,
electromagnetic transition probabilities $B_{if}(XL)$ were measured
which are squares of transition matrix elements according to
Eqs.~(\ref{eqII.3}) and~(\ref{eqII.4}).  In Ref.~\cite{Auger,Brody} 
it is argued that
the distribution of the transition matrix elements $W_{if}$ is, once more, 
of the
form~(\ref{eqIII.0}) and~(\ref{eqIII.2}). In the present section, 
however, we give a
derivation, valid for large $N$, which is well suited for the discussion of
isospin breaking in Sec.~\ref{IV}.

In our model, the initial state $|i\rangle$ and the final state
$|f\rangle$, are represented by the eigenvectors $u_i$ and $u_f$,
respectively.  The electromagnetic transition operator ${\cal O}(XL)$
is modeled by one fixed random matrix. It is not necessary to consider
an ensemble of such operators.  Thus, the transition matrix elements
$W_{if}(XL)$ are given by
\begin{eqnarray}
 W_{if}(XL) &=& u_{f}^{T} \, {\cal O}(XL) \, u_{i} 
                 \nonumber \\
            &=& \sum_{nm} u_{fn} {\cal O}_{nm}(XL) u_{im} \ .
\label{eqIIIB.1}
\end{eqnarray}
The transition probabilities read $B_{if}(XL)=|W_{if}(XL)|^2$.
Although these quantities differ from the partial widths and their
amplitudes, it is intuitively obvious that, for large $N$, the
$W_{if}(XL)$ are Gaussian distributed and that the $B_{if}(XL)$ obey a
Porter--Thomas law: Since we always consider $i\ne f$, the
$W_{if}(XL)$ are linear combinations of (products of two) independent
variables and,
therefore, the central limit theorem applies.

More precisely, the distribution of the matrix elements $W_{if}$ reads
\begin{eqnarray}
P(W_{if}) &=& \int d[u_i] P(u_i) \int d[u_f] P(u_f)
  \nonumber \\ 
    & & \qquad\qquad \delta\left(W_{if} - u_f^T{\cal O}u_i\right) \ ,
\label{a1}
\end{eqnarray}
where we suppress the argument $XL$ of $W_{if}$ and ${\cal O}$.
Since we may assume that $N$ is large, the distributions of the
components $u_{im}$ and $u_{fm}$ take the form
\begin{eqnarray}
  P(u) &=& \prod_{m=1}^N \sqrt{\frac{N}{2\pi}}
              \exp\left(-\frac{N}{2}u_m^2\right) 
  \nonumber \\ 
       &=& \sqrt{\frac{N}{2\pi}}^N
              \exp\left(-\frac{N}{2}u^Tu\right)
\label{a2}
\end{eqnarray}
with $u$ standing for either $u_i$ or $u_f$. We notice that the
integration domain for each of the eigenvector components is the
interval $[-1,+1]$. However, since we consider the large $N$ limit,
the distributions~(\ref{a2}) are so sharply peaked at the origin
that we may extend the domain of integration to the entire real axis.
Using the Fourier transform of the $\delta$ function in Eq.~(\ref{a1})
we have
\begin{eqnarray}
P(W_{if}) = \frac{1}{2\pi} \int_{-\infty}^{+\infty} dt
            \exp\left(itW_{if}\right) R(t,{\cal O}) \ .
\label{a3}
\end{eqnarray}
The Gaussian integrals absorbed in $R(t,{\cal O})$ can be done
easily,
\begin{eqnarray}
R(t,{\cal O}) &=& \sqrt{\frac{N}{2\pi}}^{2N} \int d[u_i] \int d[u_f] 
                  \nonumber \\
              & & \exp\left(-\frac{N}{2}[u_i^T \ u_f^T] 
                  \left[ 
                    \begin{array}{cc} 
                         \openone _N & it{\cal O}/N  \\
                         it{\cal O}/N  & \openone _N
                    \end{array} 
                  \right]
                  \left[ 
                    \begin{array}{c} 
                         u_i  \\
                         u_f  
                    \end{array} 
                  \right] 
                  \right) \nonumber \\
              &=& {\rm det}^{-1/2}\left(\openone _N+t^2{\cal O}^2/N^2\right)
                 \nonumber \\
              &=& \exp\left(-\frac{1}{2}
              {\rm tr}\ln\left(\openone _N+t^2{\cal O}^2/N^2\right)\right) \ .
\label{a4}
\end{eqnarray}
We are interested in large $N$. This allows us to expand the logarithm
and to keep only the first term which is $t^2{\cal O}^2/N^2$.  Thus we
have for large $N$,
\begin{eqnarray}
R(t,{\cal O}) = \exp\left(-\frac{\overline{{\cal O}^2}}{2}t^2\right)
                                            \ ,
\label{a5}
\end{eqnarray}
where the second moment of the operator is defined by
\begin{eqnarray}
\overline{{\cal O}^2} = \frac{1}{N^2} {\rm tr\,}{\cal O}^2 \ . 
\label{secmo}
\end{eqnarray}
We assume that the matrix elements ${\cal O}_{nm}(XL)$ do not depend on $N$. 
Hence, collecting everything, the Fourier transform in
Eq.~(\ref{a3}) yields
\begin{eqnarray}
  P(W_{if})= G(W_{if}, \overline{{\cal O}^{2}}^{1/2})\ , 
\label{dist}
\end{eqnarray} 
which is the expected result.

\section{Random Matrix Model --- the case of isospin breaking}
\label{IV}

In Sec.~\ref{IVA}, the model is set up for the case that isospin is
partially conserved.  In Sec.~\ref{IVB}, numerical  
simulations are described. Note that we speak of isospin breaking 
because we have in mind the experiment of Sec.~\ref{II}. The model, 
however, applies to any other quantum number that is partially conserved.

\subsection{Definition of the model}
\label{IVA}

We consider two isospin values $T=0$ and $T=1$. If isospin were
fully conserved, the Hamiltonian $H$ would be block diagonal.
The Coulomb interaction, however, destroys this symmetry.
In Ref.~\cite{Weid}, this was modeled by using random matrices
of the form
\begin {eqnarray}
  H = \left[ 
  \begin{array}{cc} 
     H(0) & 0\\
     0 & H(1) 
  \end{array}\right]
  +  \alpha  \left[
  \begin{array}{cc} 
     0 & H_{C} \\
     H_{C}^{T} & 0
  \end{array}\right], 
\label{Hamiltonian}
\end {eqnarray}
where $H(j), \ j=0,1$ are independent GOE matrices with dimension
$N_{j}$.  The total level number is $N=N_0+N_1$. The $N_0 \times N_1$
matrix $H_{C}$ accounts for the Coulomb interaction. It is real
without further symmetries and has Gaussian distributed entries. In
this model, the parameter $\alpha$ is proportional to the root mean
square Coulomb matrix element. Since $H(j)$ and $H_C$ have positive
and negative entries with equal weight, all observables can only
depend on the modulus of $\alpha$, but not on its sign. Thus, in the
sequel, we restrict ourselves to $\alpha \ge 0$. We recall that the
{\it spectral} fluctuations are measured on the scale of the local
mean level spacing $D$. Thus, the relevant parameter governing the
spectral fluctuations is
\begin{eqnarray}
\lambda=\alpha /D .
\label{23}
\end{eqnarray}

For $\alpha =0$ we have a non--interacting superposition of two
independent GOE's. In this case, the distribution of the eigenvector
components of the {\it full} matrix $H$, i.e.~including the many exact
zeros, is given by
\begin{equation}
P_{N_0N_1}(a,0) = g_0^2 P_{N_0}(a) + g_1^2 P_{N_1}(a)
                                   + 2g_0g_1 \delta(a) \ ,
\label{dist0}
\end{equation}
where we have introduced the fractional level numbers
\begin{equation}
g_0 = N_0/N \qquad {\rm and} \qquad g_1 = N_1/N \ .
\label{frac}
\end{equation}
The distributions $P_{N_j}(a), \ j=0,1$ have to be taken as either
Eq.~(\ref{eqIII.0}) or~(\ref{eqIII.1}). Obviously, the total
distribution $P_{N_0N_1}(a,0)$ is properly normalized.

As discussed in Ref.~\cite{Weid}, the variances of the distributions
for the matrix elements are chosen in such a way that the spectra of
$H(0)$ and $H(1)$ have the same length and that $H$ becomes
a full $N \times N$ GOE matrix~\cite{ERR} for $\alpha=1$ if $N_0=N_1$. 
This also means that
\begin{equation}
P_{N/2,N/2}(a,1) = P_{N}(a) \ . 
\label{dist1}
\end{equation}
For $\alpha=1$ but $N_0 \neq N_1$ there are some deviations from the
pure GOE results.

For arbitrary $\alpha$, the distribution $P_{N_0N_1}(a,\alpha)$ is not
known analytically. A qualitative model is presented in Sec.~\ref{V} (see
Eq.~(\ref{rat1})).

The transition matrix elements $W_{if}(XL)$ and the corresponding
transition probabilities $B_{if}(XL)$ depend not only on the
eigenstates $u_i$ of $H$, but also on the transition operator ${\cal
  O}(XL)$. Since the latter contains the effective charges or the
magnetic $g$ factors of proton and neutron, it causes 
isospin breaking in addition to the isospin breaking built 
into~(\ref{Hamiltonian}).  Thus, in the same isospin basis that was used in
Eq.~(\ref{Hamiltonian}), we may model the transition operator by a
matrix of the form
\begin {eqnarray}
  {\cal O} = \beta_I \left[ 
  \begin{array}{cc} 
     {\cal O}(0) & 0\\
     0 & {\cal O}(1)
  \end{array}\right] 
  +  \beta_C \left[
  \begin{array}{cc} 
     0 & {\cal O}_{C}\\
     {\cal O}_{C}^{T} & 0
  \end{array}\right] \ .
\label{Operator}
\end {eqnarray}
Here, each of the matrices ${\cal O}(j)$, ${\cal O}_{C}$ will be modeled 
by one fixed random matrix --- as described in Sec.~\ref{IVB}. 
For later convenience, ${\cal O}$ has been
written as a function of the two parameters $\beta _{I}$ and $\beta _{C}$.
However, only the ratio
\begin{eqnarray}
\beta = \beta _{C}/\beta _{I}
\label{28}
\end{eqnarray}
is important since the total strength of ${\cal O}$ drops out of the 
observable $y$ of
Eq.~(\ref{eqII.1}). For $\beta =0$, configurations with different 
isospin values are not mixed by ${\cal O}$ while for $\beta =1$ 
the mixing is maximal. For $\beta \rightarrow \infty $ the operator 
${\cal O}$ couples configurations with different isospin only.

We are interested in the distribution of the $W_{if}(XL)$ and the
$B_{if}(XL)$. According to the discussion in the previous
Sec.~\ref{III}, one is tempted to argue as follows: Since
Eq.~(\ref{eqIIIB.1}) is general and also holds in the presence of
conserved or broken isospin, the transition matrix elements
$W_{if}(XL)$ are always a linear combination of products of
independently distributed sets of variables $u_{im}$ and $u_{fn}$.
Thus, the central limit theorem should apply and we should obtain a
Gaussian distribution for the $W_{if}(XL)$. This should even be true
for any smooth distribution for the wave function components $u_{im}$
and $u_{fn}$. In other words, the specific form of these latter
distributions, which are non--trivial functions of the mixing
parameter $\alpha $, is not important and we would always expect a
Gaussian distribution for the transition matrix elements $W_{if}(XL)$.
However, this reasoning is incorrect, because it does not make use of a
specific basis. In other words: The result of this reasoning is an
orthogonally invariant distribution. Within the random matrix
model~(\ref{Hamiltonian}) one sees that this cannot be true. The
statistical properties of the Hamiltonian $H$ are not orthogonally
invariant --- except for the special case of no isospin symmetry at all.
One sees the lack of orthogonal invariance very clearly in the
distribution~(\ref{dist0}) of the eigenvector components in the case
$\alpha =0$: The $\delta $-distribution will appear only if the basis
vectors have well defined isospin. In general, the wave function components
are functions of $\alpha $. These functions parametrically depend on $N$.
This $N$-dependence
competes with a large $N$ expansion needed in the derivation of the
central limit theorem --- whose premises are therefore violated.  We
illustrate this in Sec.~\ref{IVB} by numerical examples.

\subsection{Numerical simulation}
\label{IVB}

The distributions of the reduced transition amplitudes and reduced transition 
probabilities were numerically investigated. To this end, random Hamiltonians
with the structure of Eq.~(\ref{Hamiltonian}) have been constructed. The
dimensions $N_{0}$ and $N_{1}$ were chosen to be 100 so that $H$ has the
dimension $N=200$. The elements of $H(j)$, $j=0,1$, were selected by a generator
of Gaussian random numbers such that the second moments were 
\begin{eqnarray}
H_{\mu \nu} (0) H_{\mu ^{\prime} \nu ^{\prime}} (0) &=& 
     \delta _{\mu \mu ^{\prime}} \delta _{\nu \nu ^{\prime}} + 
     \delta _{\mu \nu ^{\prime}} \delta _{\nu \mu ^{\prime}} \nonumber \\
H_{mn} (1) H_{m^{\prime} n^{\prime}} (1) &=&
     \delta _{m m^{\prime}} \delta _{n  n^{\prime}} + 
     \delta _{m n^{\prime}} \delta _{n m^{\prime}} \nonumber \\
(H_{C})_{\mu m} (H_{C})_{\mu ^{\prime} m^{\prime}} &=&
     \delta _{\mu \mu ^{\prime}} \delta _{m m^{\prime}} .
\label{30}
\end{eqnarray} 
This ensures that $H$ is a full GOE matrix for $\alpha =1$. The mean spacing
$D_{j}$ of the eigenvalues of $H(j)$ is then (in the center of the spectrum)
\begin{eqnarray}
D_{j} = \pi N_{j}^{-1/2} = 0.314 .
\label{31}
\end{eqnarray}
The mean level spacing $D$ of $H$ is 
\begin{eqnarray}
D=(D_{0}^{-1} + D_{1}^{-1})^{-1} = 0.157 .
\label{32}
\end{eqnarray}
This value changes little, when $\alpha $ is varied between 0 and 0.157. This
range of values was considered in the present numerical simulations. The
parameter $\lambda $ of Eq.~(\ref{23}) then covers the range of 
$0 \leq \lambda \leq
1.0$. Ten Hamiltonians were constructed and diagonalized for each value of
$\lambda $.

The transition operator ${\cal O}$ was constructed very much as the Hamiltonians; 
i.e.~its elements were selected by the random number generator. 
The variances were
chosen in complete analogy with Eqs.~(\ref{30}). However, for each value of 
$\beta $, one operator has been generated. The range of $0 \leq \beta \leq 1.0$
was covered.

For given parameters $\lambda $, $\beta $ the transition 
amplitudes~(\ref{eqIIIB.1}) have been calculated with the 
indices $i,f$  running over the
eigenvectors of the ten Hamiltonians. The set of $2 \cdot 10^{5}$ 
numbers $W_{if}$ with
$i > f$ forms the numerically generated data.

In a first step, we have checked whether these data follow a Gaussian
distribution. To this end one can bin the set of $W_{if}$ 
and compare the histogram
to a Gaussian; similarly one can compare the distribution of the variable $y$ of
Eq.~(\ref{eqII.1}) to the distribution~(\ref{eqII.5}) or the distribution of $z$ of
Eq.~(\ref{eqII.2a}) to $P(z)$ given in~(\ref{eqII.6}). In order to make the
comparison --- as much as possible --- independent of the parametrization of the
statistical variable, the generalized entropy
\begin{eqnarray}
S= - \int dx \,\, p(x) \,\, \ln \frac{p(x)}{P(x)} 
\label{eqIVA.3}
\end{eqnarray}
has been used to express the difference between the distribution $p$ of the data
and the ansatz $P$. This expression is independent of the parametrization $x$.
Since both, $p$ and $P$, are normalized, $S$ is never positive. It vanishes if
and only if $p \equiv P$. The data do of course not provide a continuous
probability density $p(x)$ but rather a histogram where the $k-$th bin is
centered at $x_{k}$ and contributes the probability $p(x_{k})\Delta x_{k}$.
Therefore they provide the approximation
\begin{eqnarray}
S \approx - \sum _{k} \Delta x_{k} \, p(x_{k}) \, \ln \frac{p(x_{k})}{P(x_{k})}
\label{34}
\end{eqnarray}
to expression~(\ref{eqIVA.3}). To the extent that~(\ref{34}) 
is a good approximation 
to~(\ref{eqIVA.3}), expression~(\ref{34}) is independent 
of the parametrization. This holds
in the present case to within a few times $10^{-4}$ due to the large ensemble of
numerical data. In Tab.~\ref{entropy}, the entropy~(\ref{34}) is given for a few
values of $\beta $ and $\lambda $ in the range of 
$0 \leq \beta $, $\lambda \leq 1$.
One finds that the distribution of the data does depend on the parameters
$\lambda $ and $\beta $ which govern the symmetry breaking. The entropy
approaches zero, when $\beta $ or $\lambda $ approach unity, i.e.~when the
symmetry is strongly broken. Hence, in this case the transition amplitudes
$W_{if}$ have a Gaussian distribution. This does not hold if $\beta $ and
$\lambda $ have values much smaller than 1, in which case neither the 
Hamiltonian~(\ref{Hamiltonian}) 
nor the operator~(\ref{Operator}) significantly breaks the
symmetry.
\begin{table}
\caption{Entropy for different values of the parameters $\beta $ 
  and $\lambda $. The parameter $\alpha $ is also shown.}
\begin{tabular}{ c | c | c c c c c c }
   & $\beta \rightarrow$ & 0 & 0.02 & 0.06 & 0.1 & 0.4 & 1.0 \\
   $\lambda$ & $\alpha $ $\downarrow$ & & & & & & \\
\tableline
 0 &  0     & -1.05 & -0.75 & -0.53 & -0.37 & -0.04 & -0.0003  \\
 0.012 & 0.002 & -0.54 & -0.53 & -0.40 & -0.30 & -0.04 & -0.0003 \\
 0.019 & 0.003 & -0.43 & -0.42 & -0.34 & -0.26 & -0.04 & -0.0003 \\
 0.031 & 0.005 & -0.35 & -0.34 & -0.29 & -0.23 & -0.03 & -0.0003 \\
 0.125 & 0.02  & -0.06 & -0.06 & -0.06 & -0.06 & -0.01 & -0.0003 \\
 1.0   & 0.16  & -0.0006 & -0.0006 & -0.0006 & -0.0004 & -0.0004 & -0.0004 \\
\end{tabular}
\label{entropy}
\end{table}

\begin{figure} [t!]
\centerline{\psfig{figure=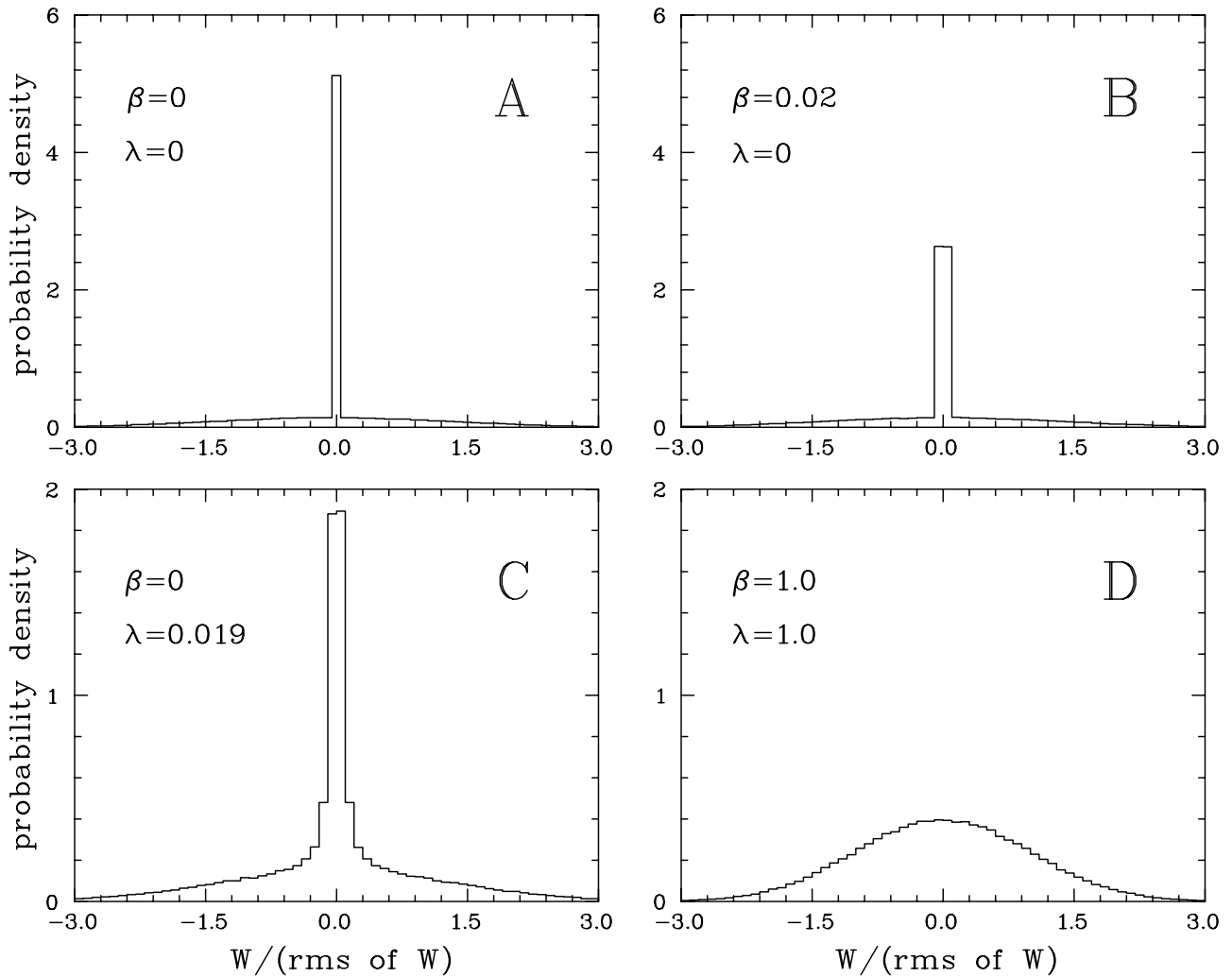,%
angle=90,width=17cm}}
\vskip 0.3cm
\caption{\protect Numerically generated distributions of transition matrix
                  elements $W_{if}$ for four different pairs of mixing parameters
                  $\beta $ and $\lambda $. The matrix elements are normalized to
                  their root mean square value. The binning of the data 
                  is the same in the four parts of the figure. The scale 
                  of the ordinate is different in the upper and lower parts.}
\label{fig1}
\end{figure}

\begin{figure} [t!]
\centerline{\psfig{figure=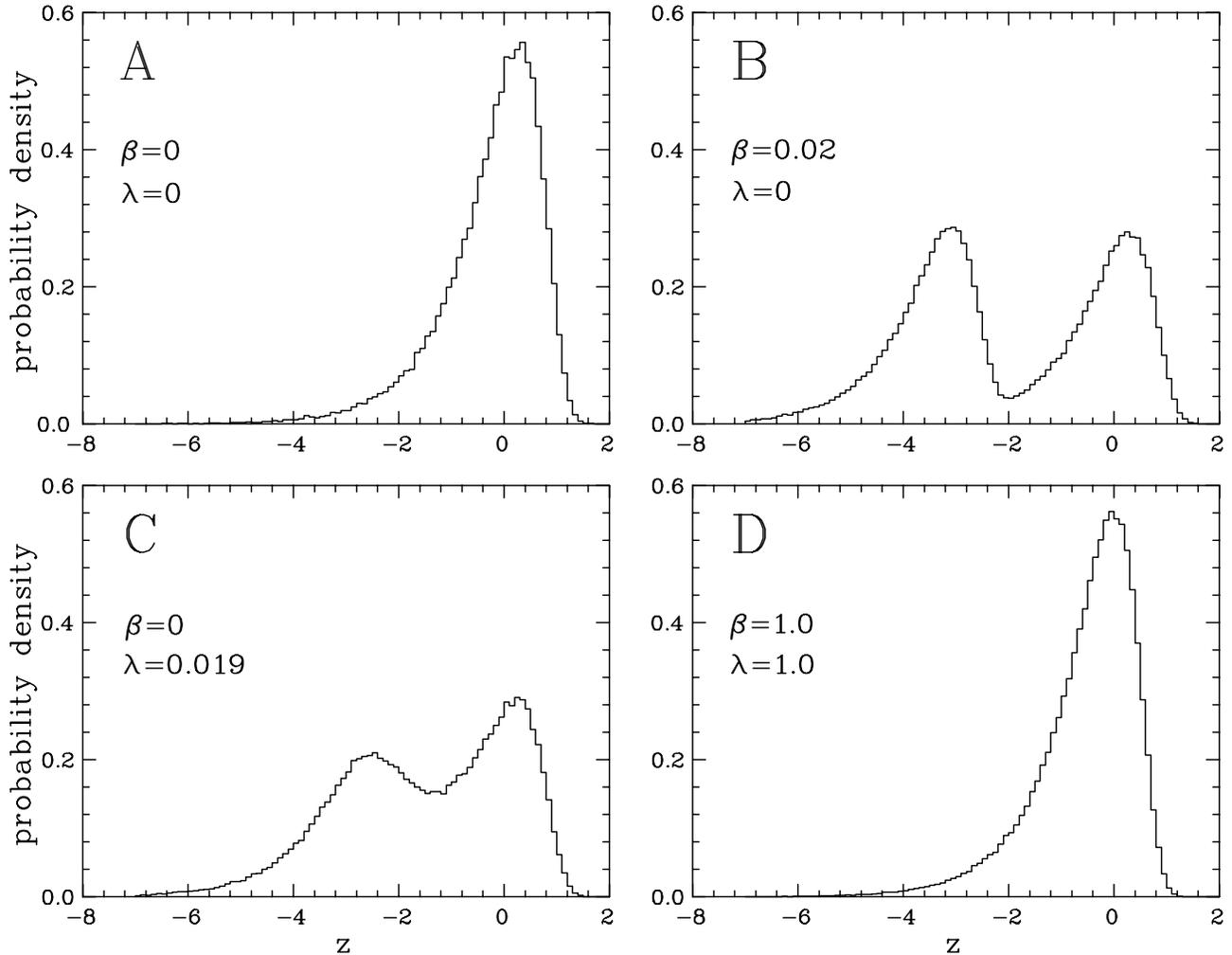,%
angle=90,width=17cm}}
\vskip 0.3cm
\caption{\protect Distributions of numerical transition probabilities
                  $B_{if}$. These are
                  the same data as in Fig.2, but reparametrized
                  in terms of the variable $z$ of Eq.~(6).}
\label{fig3}
\end{figure}

The figures~\ref{fig1} and~\ref{fig3} illustrate in which way the 
distribution of $W_{if}$ deviates from a Gaussian. Consider the case 
of $\beta =0=\lambda $, i.e.~the absence of symmetry breaking, 
displayed in Fig.~\ref{fig1}A. There is no transition between different 
symmetry classes and therefore about one half of the matrix elements 
$W_{if}$ vanishes. This leads to the narrow peak in
Fig.~\ref{fig1}A comprising one bin centered at zero. This peak is superimposed
over the distribution of the matrix elements connecting states in one and the
same symmetry class. Fig.~\ref{fig1}A thus corresponds to Eq.~(\ref{dist0}) with
$N_{0}=N_{1}$. When the symmetry is completely broken --- in Fig.~\ref{fig1}D
--- the matrix elements $W_{if}$ have a Gaussian distribution. Intermediate
situations are displayed in the parts B and C of Fig.~\ref{fig1}.

The intermediate situation is especially well visualized on Fig.~\ref{fig3},
where the distributions of Fig.~\ref{fig1} have been transformed to the
logarithmic variable
\begin{eqnarray}
z=\log _{10} \frac{W_{if}^{2}}{\overline{W_{if}^{2}}}  \ ,
\label{36}
\end{eqnarray}
which was introduced in Eq.~(\ref{eqII.2a}).

\end{multicols}

\widetext

\begin{multicols}{2}

\narrowtext

Here, the distribution displays two peaks. One of them --- centered at
$z=0$ --- stems from the matrix elements connecting states within one
symmetry class. The other one --- centered at negative values of $z$
--- is due to the matrix elements connecting states from different
symmetry classes. In case of complete symmetry breaking the two peaks
merge as in Fig.~\ref{fig3}D. This distribution corresponds to
Eq.~(\ref{eqII.6}). In the present parametrization the case of no
symmetry breaking (Fig.~\ref{fig3}A) cannot be distinguished from the
case of complete symmetry breaking (Fig.~\ref{fig3}D). This occurs
because $z \rightarrow - \infty $ when $W \rightarrow 0$; hence, the
matrix elements with the value $W=0$ are found nowhere on the
$z-$axis.

In Sec.~\ref{II}, it has been described how the secular variations of the
statistical properties of the experimental data of
Ref.~\cite{26Al} have been removed by help of the local average in
Eq.~(\ref{eqII.1}). We have applied the same
procedure to some of the present numerically generated data sets. 
Introducing
a local average $\langle B_{if} \rangle $ instead of the global average
$\overline {W_{if}^{2}} = \overline {B_{if}}$ used in
Figs.~\ref{fig1},\ref{fig3}, does not change the present results; 
i.e.~along the spectrum
of eigenstates there are no secular variations of the statistical
properties of the eigenfunctions.

The numerical simulations clearly demonstrate that there indeed is an
effect of symmetry breaking on the distributions. Hence, it is shown
that the complexity of the wave functions alone is not enough to
ensure Gaussian distributions of wave function, width and transition
amplitudes, as a naive application of the central limit theorem would
imply.

\section{Qualitative evaluation of the random matrix model with broken symmetry}
\label{V}

In the present section, we discuss the random matrix model in a
heuristic fashion. This is an extension of the Sec.~\ref{III} to the
case of broken symmetry.  Although we believe that the model of
Eqs.~(\ref{Hamiltonian},\ref{Operator}) can be completely solved
analytically, we refrain from tackling that task. As we shall see,
qualitative reasoning will provide sufficient understanding of the
statistical behavior of the observables. Thus, our analytical
derivations will be approximate and only exact in some limiting cases.
Of course, we shall compare our approximate results with the above
numerical simulations of the full model.

To obtain the distribution $P(W)$ we use the ansatz~(\ref{Operator})
for the transition operator and we need an ansatz for the
distributions of the wave functions. The latter ones --- in our model
of the eigenvector components --- are the entries of the orthogonal
matrix $U$ which diagonalizes the Hamiltonian,
\begin{eqnarray}
  U = \left[ 
  \begin{array}{cc} 
     U(0) & U(01) \\
     U(10) & U(1) 
  \end{array}\right] \ ,
\label{Umatrix}
\end{eqnarray}
where the matrices $U(j), \ j=0,1$ are $N_j\times N_j$ orthogonal and
$U(01)$ and $U(10)$ are $N_0\times N_1$ and $N_1\times N_0$ matrices,
respectively, with the condition that the total $U$ is orthogonal.
Motivated by the numerical simulations in the previous section and by
the general experience with wave function distributions, we assume
that the distributions of the entries can be approximated by
Gaussians. More precisely, we estimate that the distributions of the
$U(j)$ matrix elements are roughly given by $G(a,\kappa_{j}N^{-1/2}),
\ j=0,1$, while those of the $U(01)$ and $U(10)$ matrix elements are
roughly given by $G(a,\eta N^{-1/2})$.  Thus, we reduce the problem to
the determination of the three parameters $\kappa_0$, $\kappa_1$ and
$\eta$. It is reasonable to assume that the former two are functions
of the latter one, i.e.~we set $\kappa_j=\kappa_j(\eta)$.  This is
motivated by the fact that $\eta $ can replace the mixing parameter
$\alpha $ (or $\lambda $) and that --- apart from the level numbers
$N_{0}$, $N_{1}$ --- the distribution of the elements of $U$ is a
function of the symmetry breaking. The final task is then to find the
relation between the purely phenomenological parameter $\eta$ and the
parameter $\alpha$ which measures the root mean square symmetry
breaking matrix element and thus has a direct physical interpretation.
In other words, we eventually have to specify the function
$\eta=\eta(\alpha)$.

\end{multicols}

\widetext
\vskip -0.4cm

\begin{multicols}{2}

\narrowtext

To construct a consistent guess for the functions
$\kappa_j=\kappa_j(\eta)$, we notice that, in line with our basic
assumptions, the combined distribution for the matrix elements of the
total $U$ reads
\begin{eqnarray}
P_{N_0N_1}(a,\eta) &=& g_0^2 G\left( a,\frac{\kappa_0(\eta)}{\sqrt{N}}\right) + 
                     g_1^2 G\left( a,\frac{\kappa_1(\eta)}{\sqrt{N}}\right) 
                     \nonumber\\
                   & & \qquad \qquad + 2g_0g_1 G\left(
                   a,\frac{\eta}{\sqrt{N}}\right) \ . 
\label{rat1}
\end{eqnarray}
This form is consistent with the exact results~(\ref{dist0}) for
arbitrary $N_0$ and $N_1$ and~(\ref{dist1}) for $N_0=N_1$. We mention
in passing that we always deal with Gaussian approximations of
distributions which, in some limiting cases, are for finite level
number given by Eq.~(\ref{eqIII.0}).

Let us consider the case $\alpha=0$ which yields the
distribution~(\ref{dist0}). Comparison with Eq.~(\ref{rat1}) shows
that
\begin{equation}
\eta^2(0) = 0
\label{eta0}
\end{equation}
and
\begin{equation}
\kappa_j^2(0) = 1/g_j \ , \quad j=0,1 \ .
\label{kappa0}
\end{equation}
In the case of $\alpha=1$ and $N_0=N_1$, the exact result~(\ref{dist1})
leads to
\begin{equation}
\eta^2(1) = 1
\label{eta1}
\end{equation}
and
\begin{equation}
\kappa_j^2(1) = 1 \ , \quad j=0,1 \ .
\label{kappa1}
\end{equation}
Moreover, the functions $\kappa_j=\kappa_j(\eta)$ ought to be even in
$\eta$. This suggests the ansatz
\begin{equation}
\kappa_j^2(\eta) = \frac{1}{g_j} + 
               \left(1-\frac{1}{g_j}\right)\eta^2 \ , \quad j=0,1 \ . 
\label{kappa}
\end{equation}
It is reassuring that similar combinations of parameters do also show
up in exact calculations of the spectral correlators.

\begin{figure} [t!]
\centerline{\psfig{figure=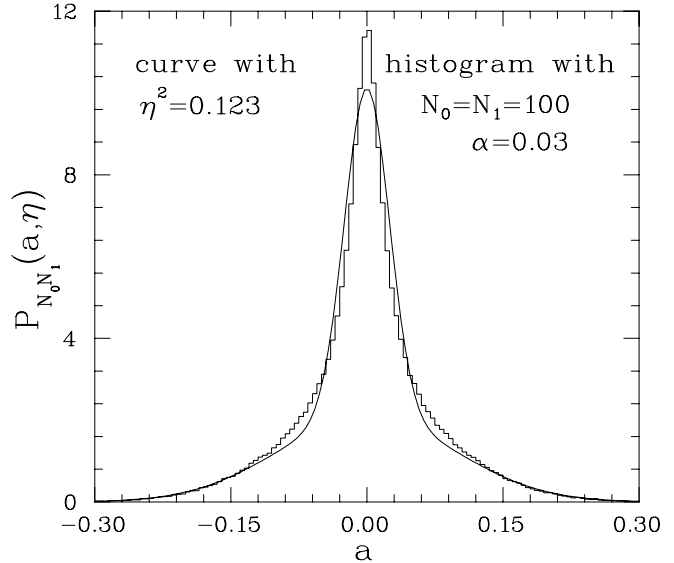,%
angle=90,width=8.6cm}}
\vskip 0.3cm
\caption{\protect The distribution of the components $a$ of the eigenfunctions
                  of $H$. The parameters of the numerical simulation are given
                  on the figure. This yields the histogram. The curve is the
                  function~(37) with the optimum $\eta $.}
\label{fig4}
\end{figure}

\begin{figure} [t!]
\centerline{\psfig{figure=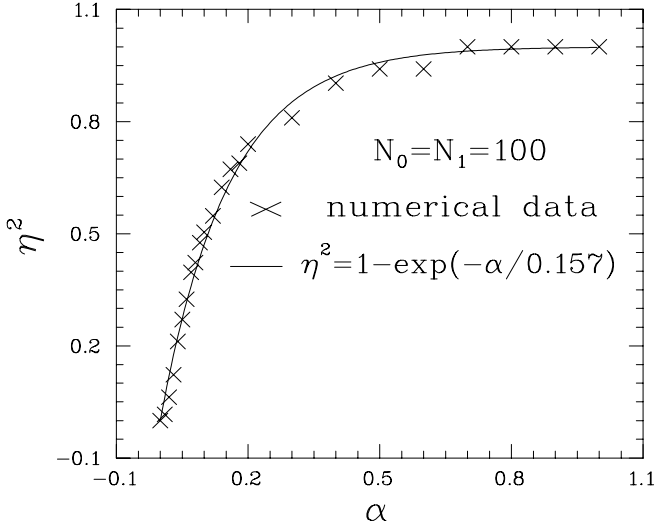,%
angle=90,width=8.6cm}}
\vskip 0.2cm
\caption{\protect The relation between $\alpha $ and $\eta ^{2}$ for
                  $N_{0}=N_{1}$. The crosses are from the numerical simulation.
                  The curve is a phenomenological representation of the
                  numerical results.}
\label{fig5}
\end{figure}

A reasonable form for the function
$\eta=\eta(\alpha)$ remains to be given. The function will parametrically
depend on $g_0$ and $g_1$.  We construct it by
comparing the ansatz (\ref{rat1}) with the distribution of the coefficients
$u_{im}$ of the eigenfunctions $u_{i}$ from the numerical simulation of
Sec.~\ref{IVB}. For each value of $\lambda =\alpha /D$, the optimum $\eta $ was determined
by maximizing the entropy~(\ref{34}).
We have not attempted to assign errors to the values of $\eta $ thus found.
Instead we have convinced ourselves by inspection that the ansatz~(\ref{rat1})
reproduces the numerical distribution reasonably well. A typical example is
given by Fig.~\ref{fig4}. In this way the function $\eta = \eta (\alpha )$ was
generated numerically. For the case of $N_{0}=N_{1}$ is is displayed on
Fig.~\ref{fig5} together with the analytical expression 
\begin{eqnarray}
\eta ^{2}(\alpha) = 1- \exp (- \alpha/0.157 )
\label{42a}
\end{eqnarray}
that well approximates it. The numerical constant in the exponential
is here equal to the level distance of Eq.~(\ref{32}). 
We emphasize, however, that the relation~(\ref{42a}) 
is purely phenomenological. In Sec.~\ref{VI}, we shall see that~(\ref{42a})
can successfully be applied to the case of $^{26}$Al, where $D$ is different
from~(\ref{32}), see Eq.~(\ref{73}) below, but the difference is not too
large. Thus it seems that the numerical constant is indeed related to the
level distance. This quantity shows a significant $N$ dependence, see
Eq.~(\ref{31}). It is at this point where the
premises of the central limit theorem are violated for the
calculations of the distributions of the transition matrix elements.
They cannot be a single Gaussian for all values of the parameters.

Having fixed all parameters in our qualitative model for the
distribution of the eigenvector components, we can now calculate the
distributions for the transition matrix elements
$W_{if}=u_f^T{\cal O}u_i$. We emphasize that this involves, apart from
the large $N$ limit, no further approximation. We consider $i \neq f$,
such that the initial and final states $u_i$ and $u_f$ 
are always different columns of the matrix $U$ in Eq.~(\ref{Umatrix}).
Due to isospin breaking, we have to distinguish four structurally
different possibilities: in case (M00) both, $u_i$ and $u_f$, are
among the first $N_0$ columns. We write
\begin{displaymath}
 u_i = \left[ 
  \begin{array}{c} 
     u_i(0) \\
     u_i(10)  
  \end{array}\right] 
\quad {\rm and} \quad
 u_f = \left[ 
  \begin{array}{c} 
     u_f(0) \\
     u_f(10)  
  \end{array}\right] \ . 
\qquad\quad {\rm (M00)}
\end{displaymath}
In case (M11), both are from the last $N_1$ columns,
\begin{displaymath}
 u_i = \left[ 
  \begin{array}{c} 
     u_i(01) \\
     u_i(1)  
  \end{array}\right] 
\quad {\rm and} \quad
 u_f = \left[ 
  \begin{array}{c} 
     u_f(01) \\
     u_f(1)  
  \end{array}\right] \ . 
\qquad\quad {\rm (M11)}
\end{displaymath}
In the cases (M01) and (M10), $u_i$ and $u_f$ are from different parts
in $U$, we have 
\begin{displaymath}
 u_i = \left[ 
  \begin{array}{c} 
     u_i(0) \\
     u_i(10)  
  \end{array}\right] 
\quad {\rm and} \quad
 u_f = \left[ 
  \begin{array}{c} 
     u_f(01) \\
     u_f(1)  
  \end{array}\right] \ , 
\qquad\quad {\rm (M01)}
\end{displaymath}
and 
\begin{displaymath}
 u_i = \left[ 
  \begin{array}{c} 
     u_i(01) \\
     u_i(1)  
  \end{array}\right] 
\quad {\rm and} \quad
 u_f = \left[ 
  \begin{array}{c} 
     u_f(0) \\
     u_f(10)  
  \end{array}\right] \ . 
\qquad\quad {\rm (M10)}
\end{displaymath}
Thus, for the case (M$jj^\prime$), $j,j^\prime=0,1$ we will obtain a
distribution $P_{jj^\prime}(W_{if})$. Because of the symmetries ${\cal
  O}^T={\cal O}$ and $W_{fi}=W_{if}$, the cases (M01) and (M10) must
yield identical distributions.

We proceed by generalizing the evaluation of the integral~(\ref{a1}) 
in Sec.~\ref{IIIB}. Again, the distribution 
\begin{eqnarray}
P_{jj^\prime}(W_{if}) &=& \int d[u_i] P(u_i) \int d[u_f] P(u_f)
  \nonumber \\ 
    & & \qquad\qquad \delta\left(W_{if} - u_f^T{\cal O}u_i\right) 
\label{iso1}
\end{eqnarray}
is rewritten in the form
\begin{eqnarray}
P_{jj^\prime}(W_{if}) = \frac{1}{2\pi} \int_{-\infty}^{+\infty} dt
            \exp\left(itW_{if}\right) R_{jj^\prime}(t,{\cal O}) ,
\label{iso2}
\end{eqnarray}
where the main difficulty is the calculation of the function
\begin{eqnarray}
R_{jj^\prime}(t,{\cal O}) &=& \int d[u_i] P(u_i) \int d[u_f] P(u_f) 
                       \nonumber\\
              & & \qquad\qquad 
                     \exp\left(-itu_f^T{\cal O}u_i\right) .
\label{iso3}
\end{eqnarray}

In the case (M00), the distributions of the eigenvectors are according
to our ansatz 
\begin{eqnarray}
  P(u_i) &=& \sqrt{\frac{N}{2\pi\kappa_0^2}}^{N_0}
              \exp\left(-\frac{N}{2\kappa_0^2}u_i^T(0)u_i(0)\right)
  \nonumber \\ 
       & & \quad \times \sqrt{\frac{N}{2\pi\eta ^{2}}}^{N_1}
              \exp\left(-\frac{N}{2\eta ^{2}}u_i^T(10)u_i(10)\right)
\label{iso4}
\end{eqnarray}
for the initial state and 
\begin{eqnarray}
  P(u_f) &=& \sqrt{\frac{N}{2\pi\kappa_0^2}}^{N_0}
              \exp\left(-\frac{N}{2\kappa_0^2}u_f^T(0)u_f(0)\right)
  \nonumber \\ 
       & & \quad \times \sqrt{\frac{N}{2\pi\eta^2}}^{N_1}
              \exp\left(-\frac{N}{2\eta^2}u_f^T(10)u_f(10)\right)
\label{iso5}
\end{eqnarray}
for the final state. The case (M11) is treated similarly. The case (M01) has
a different structure: The initial state distribution is as in
Eq.~(\ref{iso4}), but the final state distribution reads
\begin{eqnarray}
  P(u_f) &=& \sqrt{\frac{N}{2\pi\eta^2}}^{N_0}
              \exp\left(-\frac{N}{2\eta ^{2}}u_f^T(01)u_f(01)\right)
  \nonumber \\ 
       & & \quad \times \sqrt{\frac{N}{2\pi\kappa_1^2}}^{N_1}
           \exp\left(-\frac{N}{2\kappa_1^2}u_f^T(1)u_f(1)\right) \ .
\label{iso6}
\end{eqnarray}
However, one sees that in every case we may cast Eq.~(\ref{iso3})
into the form
\begin{eqnarray}
R_{jj^\prime}(t,{\cal O}) &=& C_{jj^\prime} \int d[u_i] \int d[u_f] 
                       \nonumber\\
  & & \qquad \exp\left(-\frac{N}{2}[u_i^T \ u_f^T]
                    \, {\cal K}_{jj^\prime}
           \left[ 
              \begin{array}{c} 
                u_i \\
                u_f
              \end{array}
           \right]
          \right)
                 \nonumber \\
      &=& C_{jj^\prime}^\prime \, {\rm det}^{-1/2}{\cal K}_{jj^\prime} .
\label{iso7}
\end{eqnarray}
Here, $C_{jj^\prime}$ and $C_{jj^\prime}^\prime$ collect all
normalization constants.

It is straightforward to write down the matrices ${\cal
  K}_{jj^\prime}$. We define the diagonal matrices
\begin{eqnarray}
D_0 &=& {\rm diag\,}\left(\frac{1}{\kappa_0^2}\openone _{N_0},
                          \frac{1}{\eta^2}\openone _{N_1}\right)
  \nonumber \\ 
D_1 &=& {\rm diag\,}\left(\frac{1}{\eta^2}\openone _{N_0},
                          \frac{1}{\kappa_1^2}\openone _{N_1}\right)
\label{iso8}
\end{eqnarray}
and find 
\begin{eqnarray}
{\cal K}_{jj^\prime} &=& \left[ 
  \begin{array}{cc} 
     D_j           & it{\cal O}/N \\
     it{\cal O}/N  & D_{j^\prime}
  \end{array}\right]  
\label{iso9}
\end{eqnarray}
in the four cases. An easy calculation yields
\begin{eqnarray}
R_{jj^\prime}(t,{\cal O}) &=& 
        {\rm det}^{-1/2}\left(\openone _N+\frac{t^2}{N^2}D_j^{-1}{\cal O}
                      D_{j^\prime}^{-1}{\cal O}\right)
                       \nonumber\\
  &=& \exp\left(-\frac{1}{2}
         {\rm tr\,}\ln\left(\openone _N+\frac{t^2}{N^2}D_j^{-1}{\cal O}
                      D_{j^\prime}^{-1}{\cal O}\right)\right) 
\label{iso10}
\end{eqnarray}
which leads in the large $N$ limit to
\begin{eqnarray}
R_{jj^\prime}(t,{\cal O}) = 
      \exp\left(-\frac{\overline{{\cal O}_{jj^\prime}^2}}{2}t^2\right) \ .
\label{iso11}
\end{eqnarray}
The effective second moment $\overline{{\cal O}_{jj^{\prime}}^{2}}$ 
of the operator ${\cal O}$ is defined by
\begin{eqnarray}
\overline{{\cal O}_{jj^\prime}^2} = \frac{1}{N^2} 
                      {\rm tr\,}D_j^{-1}{\cal O}
                       D_{j^\prime}^{-1}{\cal O} \ .
\label{iso12}
\end{eqnarray}
Hence, by performing the Fourier transform in Eq.~(\ref{iso2}), we arrive at
\begin{eqnarray}
P_{jj^{\prime}}(W_{if}) = G(W_{if},\overline{{\cal O}_{jj^{\prime}}^2}^{1/2}) \ ,
\label{iso13}
\end{eqnarray} 
i.e.~at a Gaussian in all four cases. The Gaussian form does not come as a
surprise. The crucial and non--trivial result of this calculation
is the explicit form of the effective second moments~(\ref{iso12}).
Denoting the second moments of the blocks in the
operator~(\ref{Operator}) by
\begin{eqnarray}
\overline{{\cal O}^2(j)} &=& \frac{1}{N^2} 
                      {\rm tr\,}{\cal O}^2(j) \ , \quad j=0,1
                                 \nonumber\\
\overline{{\cal O}_C^T{\cal O}_C} &=& \frac{1}{N^2} 
                      {\rm tr\,}{\cal O}_C^T{\cal O}_C \ ,
\label{iso14}
\end{eqnarray}
we find after a simple computation in the four cases 
\begin{eqnarray}
\overline{{\cal O}_{00}^2} &=& 
  \kappa_0^4\beta_I^2 \overline{{\cal O}^2(0)} + 
  \eta^4\beta_I^2 \overline{{\cal O}^2(1)} + 
  2\kappa_0^2\eta^2\beta_C^2\overline{{\cal O}_C^T{\cal O}_C}
                                 \nonumber\\
\overline{{\cal O}_{11}^2} &=& 
  \eta^4\beta_I^2 \overline{{\cal O}^2(0)} + 
  \kappa_1^4\beta_I^2 \overline{{\cal O}^2(1)} + 
  2\kappa_1^2\eta^2\beta_C^2\overline{{\cal O}_C^T{\cal O}_C}
                                 \nonumber\\
\overline{{\cal O}_{01}^2} &=& 
  \kappa_0^2\eta^2\beta_I^2 \overline{{\cal O}^2(0)} + 
  \kappa_1^2\eta^2\beta_I^2 \overline{{\cal O}^2(1)} +
         \nonumber\\ 
  & & \qquad\qquad 
      \left(\kappa_0^2\kappa_1^2+\eta^4\right)
              \beta_C^2\overline{{\cal O}_C^T{\cal O}_C}
                                 \nonumber\\
\overline{{\cal O}_{10}^2} &=& \overline{{\cal O}_{01}^2} \ .
\label{iso15}
\end{eqnarray}

The distributions $P_{jj^\prime}(W_{if})$ of Eq.~(\ref{iso13}) 
describe the cases
(M$jj^\prime$). These were defined by assuming that the structure of
the initial and final states, i.e.~their place of origin in the
matrix~(\ref{Umatrix}) is known. In a set of experimental data, this
information is usually lacking. Thus, the full distribution must be a
proper superposition of the functions $P_{jj^\prime}(W_{if})$. The
weights in this superposition can be easily constructed: In the case
(M00), there are $N_0$ states and therefore $N_0(N_0-1)$ transitions.
We recall that we exclude recoupling transitions with $i=f$.  Since,
altogether, there are $N$ states and $N(N-1)$ transitions, the weight
of the case (M00) is $N_0(N_0-1)/N(N-1)$, which yields $g_0^2$ 
for large level numbers. In the case (M11), one obviously finds $g_1^2$.
Similarly, we find $N_0N_1/N(N-1)$, i.e.~$g_0g_1$ for large
level numbers, in both of the cases (M01) and (M10). Thus,
collecting everything we eventually end up with
\begin{eqnarray}
P(W_{if}) &=& g_0^2 G(W_{if},\overline{{\cal O}_{00}^{2}} ^{1/2}) + 
                     g_1^2 G(W_{if},\overline{{\cal O}_{11}^{2}} ^{1/2}) 
                     \nonumber\\
                   & & \qquad \qquad + 2g_0g_1 
                   G(W_{if},\overline{{\cal O}_{01}^{2}} ^{1/2}) , 
\label{rat2}
\end{eqnarray}
where the individual variances are explicitly given in
Eq.~(\ref{iso15}).

The choice of the transition operator ${\cal O}$ in Sec.~\ref{IVB} implies that
the second moments of the blocks in~(\ref{Operator}) are (for large $N_{j}$)
\begin{eqnarray}
\overline{{\cal O}^{2}(j)} = N_{j}^{2}/N^{2} = g_{j}^{2}, \,\,\, j=0,1
\label{58}
\end{eqnarray}
and
\begin{eqnarray}
\overline{{\cal O}_C^T{\cal O}_C} = \frac{N_{0}N_{1}}{N^{2}} = g_{0}g_{1} .
\label{59}
\end{eqnarray}

Inserting this and the definition~(\ref{kappa}) of the functions 
$\kappa _{j} (\eta )$ into~(\ref{iso15}) leads to
\begin{eqnarray}
\overline{{\cal O}_{00}^2} &=& 
  \left[ (1-g_{1} \eta ^{2} )^{2} + \eta ^{4} g_{1}^{2} \right] \beta _{I} ^{2}
  + 2\eta ^{2} (1-g_{1} \eta ^{2} ) g_{1} \beta _{C}^{2}
                                 \nonumber\\
\overline{{\cal O}_{11}^2} &=& 
  \left[ (1-g_{0} \eta ^{2} )^{2} + \eta ^{4} g_{0}^{2} \right] \beta _{I} ^{2}
  + 2\eta ^{2} (1-g_{0} \eta ^{2} ) g_{0} \beta _{C}^{2}
                                 \nonumber\\
\overline{{\cal O}_{01}^2} &=& 
  (1 - 2g_{0} g_{1} \eta ^{2})\eta ^{2} \beta _{I}^{2} + (1- \eta ^{2} + 2
  g_{0}g_{1} \eta ^{4})\beta _{C}^{2} \ .
         \nonumber\\ 
\label{A}
\end{eqnarray}

Let us specialize these results to the case of the numerical simulation in
Sec.~\ref{IVB} and see whether both are in qualitative agreement.

The numerical simulation has been performed with $N_{0}=N_{1}$, whence,
\begin{eqnarray}
g_{0}=g_{1} = \frac{1}{2} .
\label{60}
\end{eqnarray}

Since the total strength of the operator ${\cal O}$ is immaterial --- see the
discussion of Eq.~(\ref{28}) --- one can set $\beta _{I} = 1$ which makes 
$\beta _{C} = \beta $. All this turns Eqs.~(\ref{A}) into 
the more transparent expressions
\begin{eqnarray}
\overline{{\cal O}_{00}^{2}} &=& 1-\left( \eta ^{2} - \frac{1}{2} \eta ^{4}
                                 \right) (1-\beta ^{2}) 
                                 \nonumber \\
\overline{{\cal O}_{11}^{2}} &=& \overline{{\cal O}_{00}^{2}} \nonumber \\
\overline{{\cal O}_{01}^{2}} &=& 1-\left( 1 - \eta ^{2} + \frac{1}{2} \eta ^{4}
                                 \right) (1-\beta ^{2}) 
                                 \nonumber \\ 
\overline{{\cal O}_{10}^{2}} &=& \overline{{\cal O}_{01}^{2}} .
\label{62}
\end{eqnarray}

If $\beta =1$, the operator~(\ref{Operator}) completely mixes configurations
with different isospin. In this case, Eqs.~(\ref{62}) yield
\begin{eqnarray}
\overline{{\cal O}_{00}^{2}} &=& \overline{{\cal O}_{11}^{2}} =  
\overline{{\cal O}_{01}^{2}} = \overline{{\cal O}_{10}^{2}} =1 \ .
\label{64}
\end{eqnarray}
The distribution~(\ref{rat2}) of $W_{if}$ then becomes
\begin{eqnarray}
P(W_{if}) = G(W_{if},1),
\label{65}
\end{eqnarray}
i.e.~a Gaussian for all $\eta $ and, hence, for all $\alpha $.  This
qualitatively agrees with the results of the numerical simulation, see
Tab.~\ref{entropy}.

If $\alpha =1$ (and $N_{0}=N_{1}$), the
Hamiltonian~(\ref{Hamiltonian}) is a full GOE matrix and its
eigenstates have no isospin symmetry. According to Eq.~(\ref{eta1}),
one then has $\eta ^{2}=1$, Eqs.~(\ref{62}) yield
\begin{eqnarray}
\overline{{\cal O}_{00}^{2}} &=& \overline{{\cal O}_{11}^{2}} = 
\overline{{\cal O}_{01}^{2}} = \overline{{\cal O}_{10}^{2}} \nonumber \\
&=& \frac{1}{2}(1+\beta ^{2}) \ ,
\label{66}
\end{eqnarray}
and the distribution~(\ref{rat2}) becomes
\begin{eqnarray}
P(W_{if}) = G\left(W_{if}, \left[ \frac{1}{2}(1+\beta ^{2})\right] ^{1/2}\right),
\label{67}
\end{eqnarray}
i.e.~a Gaussian for all $\beta $. Again this qualitatively agrees 
with the results of the numerical simulation, see Tab.~\ref{entropy}.

If neither the Hamiltonian~(\ref{Hamiltonian}) nor the operator~(\ref{Operator})
mixes isospin, i.e.~if $\beta $ and $\eta $ are zero, Eqs.~(\ref{62}) give
\begin{eqnarray}
\overline{{\cal O}_{00}^{2}} &=& \overline{{\cal O}_{11}^{2}} = 1
                                 \nonumber \\
\overline{{\cal O}_{01}^{2}} &=& \overline{{\cal O}_{10}^{2}} = 0.
\label{68}
\end{eqnarray}
In this limit, the distribution~(\ref{rat2}) takes the form
\begin{eqnarray}
P(W_{if}) = \frac{1}{2} (G(W_{if},1) + \delta (W_{if})) .
\label{69}
\end{eqnarray}
This says that half of the transition matrix elements is zero --- namely the
ones, where $i$ and $f$ have different isospin. The other half is Gaussian
distributed. This result is expected and agrees with the numerical simulation,
see Tab.~\ref{entropy} and Fig.~\ref{fig1}A.

\begin{figure} [t!]
\centerline{\psfig{figure=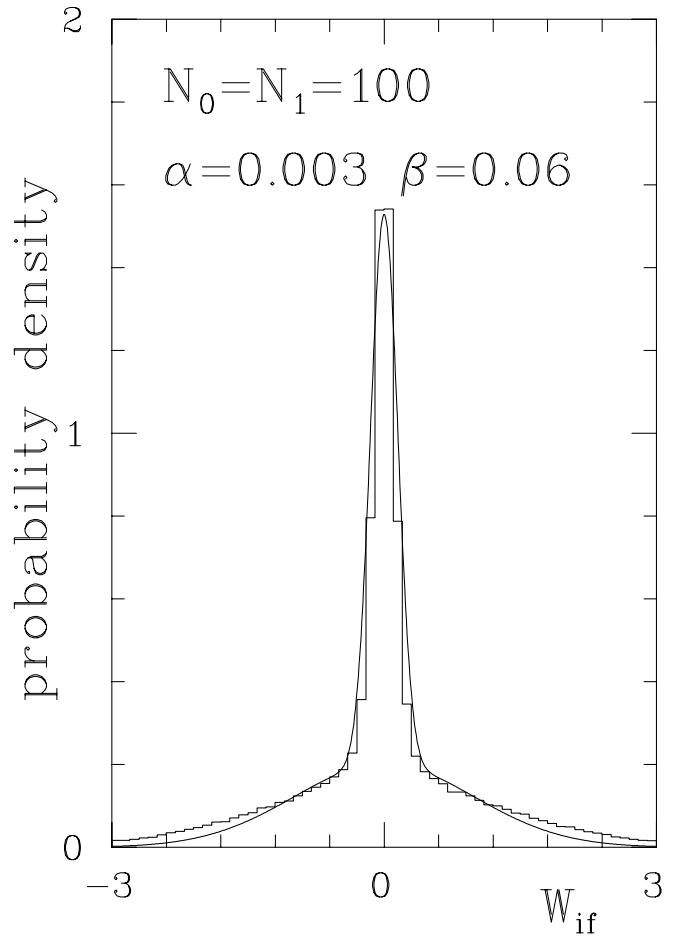,width=8.6cm}}%
\vskip 0.2cm
\caption{\protect A distribution of the transition matrix elements  $W_{if}$.
                  The histogram is due to the numerical simulation described in
                  Sec.~IVB. The curve is the qualitative 
                  model~(59) with the parameters~(71).}
\label{fig6}
\end{figure}

In Fig.~\ref{fig6}, the distribution of $W_{if}$ is displayed for the
mixing parameters $\alpha =0.003$ and $\beta =0.06$.  The histogram is
the result of the numerical simulation in Sec.~\ref{IVB}.  The curve
is the qualitative model~(\ref{rat2}). The parameters follow from
Eq.~(\ref{42a}) which yields $\eta ^{2}=0.0189$ and from
Eqs.~(\ref{62}) which give
\begin{eqnarray}
\overline{{\cal O}_{00}^{2}} &=& \overline{{\cal O}_{11}^{2}} = 0.981 \ ,
                                 \nonumber \\
\overline{{\cal O}_{01}^{2}} &=& \overline{{\cal O}_{10}^{2}} = 0.0223 \ .
\label{70}
\end{eqnarray}
This illustrates that the qualitative model reproduces the essential features 
of the numerical data although it does not describe them quantitatively.

\section{Interpretation of the experimental data}
\label{VI}

In Sec.\ref{VIA}, the experimental data described in Sec.~\ref{II} are 
compared to the prediction of the qualitative solution~(\ref{iso15}) of the 
model of Sec.~\ref{IVA}. In
Sec.~\ref{VIB}, we discuss the effect that symmetries other than isospin 
have on the data. Parity conservation is used as an example.

\subsection{Isospin breaking in $^{26}$Al}
\label{VIA}
The results of Secs.~\ref{IV} and~\ref{V} suggest that the distribution of the
reduced electromagnetic transition probabilities --- displayed in
Fig.~\ref{fig26Al} --- does carry a signature of isospin breaking. In order to
verify this, we compare in the present section the random matrix model to the
experimental data.

The parameters of the random matrix model are chosen as follows. The dimensions
$N_{0}$, $N_{1}$ of the blocks in the Hamiltonian~(\ref{Hamiltonian}) are not
equal. Rather one has
\begin{eqnarray}
N_{0}/N_{1} =3
\label{72}
\end{eqnarray}
which approximately corresponds to the ratio of the densities of states with
$T=0$ and $T=1$ in $^{26}$Al. This entails
\begin{eqnarray}
g_{0} = \frac{3}{4} , \,\,\, g_{1} = \frac{1}{4}.
\label{77}
\end{eqnarray}

\begin{figure} [hbt]
\centerline{\psfig{figure=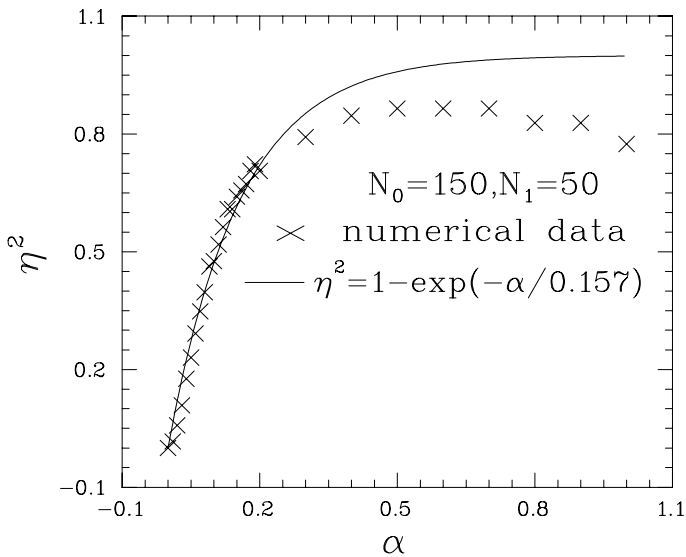,%
width=8.6cm,bblly=3.5cm,bburx=19.4cm,bbllx=1cm,bbury=24.4cm,angle=90}}
\caption{\protect The relation between $\alpha $ and $\eta ^{2}$ for 
                  $N_{0} \neq N_{1}$.}
\label{fig7}
\end{figure}

The parameter which quantifies isospin breaking in the Hamiltonian was
set to $\alpha = 0.028$. This value has been taken from the analysis
of the eigenvalue statistics in Ref.~\cite{Weid}. We must convert it
to the parameter $\eta $ of the qualitative model of Sec.~\ref{V}.
However, it is so far not clear whether Eq.~(\ref{42a}) applies
because that function is based on a numerical simulation with
$N_{0}=N_{1}$ which is different from~(\ref{72}). We have therefore
performed another numerical simulation of the distribution of the
eigenvector coefficients $a$. This simulation was done with
$N_{0}=150$ and $N_{1}=50$. For each $\alpha $, the
function~(\ref{rat1}) was again compared to the numerical distribution
of $a$ and the optimum $\eta $ was determined. This yielded the
crosses on Fig.~\ref{fig7}. It turns out that the function~(\ref{42a})
again reproduces the numerical data if one considers only the range of
$\lambda=\alpha /D \lesssim 1$. Note that in the case at hand, the
level distance is
\begin{equation}
D=0.192\ .
\label{73}
\end{equation}
For $\lambda=\alpha /D >1$ the function~(\ref{42a}) does not reproduce
the data of Fig.~\ref{fig7} because the
Hamiltonian~(\ref{Hamiltonian}) does not become a GOE matrix for any
value of $\alpha $ if $N_{0} \neq N_{1}$. Due to Eq.~(\ref{73}) the
numerical constant in~(\ref{42a}) is no longer equal to the level
distance.

One finds
\begin{eqnarray}
\eta ^{2} = 0.163
\label{74}
\end{eqnarray}
for the present case of $^{26}$Al.

Due to the fact that $^{26}$Al is a mirror nucleus, i.e.~it has the isospin
projection $T_{z}=0$, the transition operator is uniquely determined by its
isospin structure.

If ${\cal O}$ is isoscalar it can by definition not connect states with
different isospin. The parameter $\beta _{C}$ in~(\ref{Operator}) is then zero
and, of course, $\beta =0$. If the operator is isovectorial it cannot
connect states with the same isospin. This is obvious for states having $T=0$.
For states having $T=1$, this follows from the symmetry properties of ${\cal O}$
in isospin space. See Chap.~9 of Ref.~\cite{BG}, specially the discussion
following Eq.~(9.107). The parameter $\beta _{I}$ in~(\ref{Operator}) is then
zero. We evaluate the distribution~(\ref{rat2}) for these two cases.

\begin{figure} [hbt]
\centerline{\psfig{figure=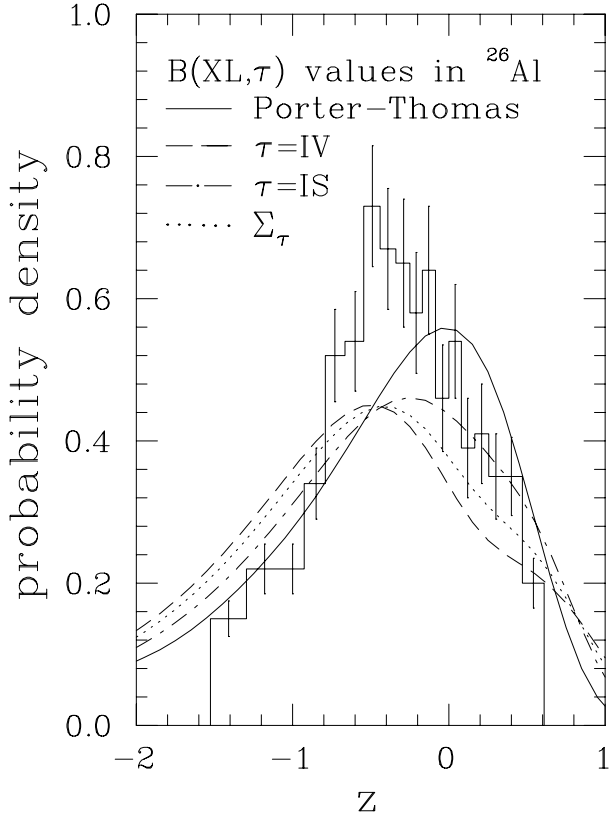,%
width=8.6cm,bblly=1.5cm,bburx=19.4cm,bbllx=1cm,bbury=24.4cm}}
\caption{\protect The distribution of reduced transition probabilities in
                  $^{26}$Al. The experimental histogram is the same as in
                  \protect Fig.~1. The full curve is the
                  Porter-Thomas distribution~(9). The remaining
                  curves have been obtained with the qualitative
                  model~(59). The dashed curve --- labeled $\tau=IV$ --- has
		  been obtained with the isovector transition operator; the
		  dot-dashed curve --- labeled $\tau = IS$ --- is from the
		  isoscalar transition operator; the dotted curve --- labeled
		  $\Sigma _{\tau }$ --- is the weighted sum of both.}
\label{fig8}
\end{figure}

Let the operator ${\cal O}$ be isoscalar, whence $\beta _{C}=0$. We
set $\beta _{I} =1$, make use of Eqs.~(\ref{77},\ref{74}), and obtain
from Eqs.~(\ref{A})
\begin{eqnarray}
\overline{{\cal O}_{00}^{2}} &=& 0.922 \nonumber \\
\overline{{\cal O}_{11}^{2}} &=& 0.785 \nonumber \\
\overline{{\cal O}_{01}^{2}} &=& \overline{{\cal O}_{10}^{2}} = 0.153\,\, .
\label{78}
\end{eqnarray}
This gives the distribution~(\ref{rat2}) of the isoscalar
matrix elements $W_{if}$. It is transformed to the variable
\begin{eqnarray}
z=\log _{10} \frac{W_{if}^{2}}{\overline{W_{if}^{2}}} \ .
\label{79}
\end{eqnarray}
The result is the dot-dashed curve on Fig.~\ref{fig8}.

Now let the operator ${\cal O}$ be isovectorial, 
whence $\beta _{I} =0$. We choose $\beta
_{C} =1$, use Eqs.~(\ref{74},\ref{77}) and obtain from Eq.~(\ref{A})
\begin{eqnarray}
\overline{{\cal O}_{00}^{2}} &=& 0.0782 \nonumber \\
\overline{{\cal O}_{11}^{2}} &=& 0.235 \nonumber \\
\overline{{\cal O}_{01}^{2}} &=& 0.847 \,\, .
\label{82}
\end{eqnarray}
This gives the distribution~(\ref{rat2}) of the isovectorial
matrix elements $W_{if}$. Again this distribution is transformed to the
variable~(\ref{79}). The result is the dashed curve on Fig.~\ref{fig8}.

One sees that both, the distribution with $\tau = IV$ and the one with $\tau
= IS$, are shifted with respect to the Porter-Thomas
distribution~(\ref{eqII.6}). This is a consequence of the appearance of
Gaussian distributions with different widths in Eq.~(\ref{rat2}). The shoulder
on the r.h.s.~of the isovector distribution in Fig.~\ref{fig8} even hints to
the appearance of two peaks as in the numerical simulations on
Fig.~\ref{fig3}B and C. However, in $^{26}$Al, isospin mixing is 
too large (namely $\lambda =0.146$) as
to let the peaks separate. Remember that it is the class of small matrix
elements that produces the peak which is centered at negative values of $z$.
The class of large matrix elements leads to a peak which is centered close to
$z=0$. The smaller the relative strength of the small matrix elements is, 
the more clearly the peaks separate. 

There is a factor other than $\alpha $ (or
$\eta ^{2}$) which determines the shape of the distributions on
Fig.~\ref{fig8}. For the isoscalar problem the class of the large matrix
elements is the one with variance $\overline {O_{00}^{2}}$, see
Eqs.~(\ref{78}). It has the largest weight in the distribution, namely
$g_{0}^{2}$, see~(\ref{rat2},\ref{77}). Therefore the distribution is shifted
only slightly to negative values of $z$. For the isovector problem the class
of the small matrix elements is the one with variance $\overline
{O_{00}^{2}}$, see Eqs.~(\ref{82}), which has the largest weight. Therefore
the distribution is shifted more strongly to negative values of $z$. This
makes the isovector matrix elements more sensitive to isospin breaking than
the isoscalar matrix elements.

The distributions for the isovector and isoscalar cases have been summed with
the weights 530 and 343, respectively. These are the numbers of observed
transitions, see Ref.~\cite{26Al}. The result is the dotted curve on
Fig.~\ref{fig8}. It is shifted with respect to the Porter-Thomas 
distribution
by about the same amount as the experimental data. However, the data are more
strongly  peaked around $z \approx -0.5$ than the present theory. This
discrepancy must be attributed --- at least partially --- to the way the
detection thresholds have been taken into account, see Eq.~(\ref{8a}): The
percentage of undetected transitions may have been underestimated because its
estimation was based on the Porter-Thomas distribution. The distributions
that include isospin violation predict a larger probability of events outside
the range of the experimental data than does the Porter-Thomas distribution.

We conclude that the present investigation supports the conjecture of
Ref.~\cite{26Al} which attributes the discrepancy between the Porter-Thomas
distribution and the experiment to partial isospin breaking in $^{26}$Al.

\subsection{Other symmetries of the system: parity}
\label{VIB}
The nucleus $^{26}$Al has more than only isospin symmetry. Especially there are
quantum numbers which are exactly conserved or nearly so. For definiteness let
us discuss parity conservation. Should it be included into all the above
considerations that have focussed on isospin conservation alone?

The matrix model of Sec.~\ref{IVA} can be extended to take parity into account.
If the system has two possible values of isospin $T=0,1$ (isospin being broken
to some extent) and parity $\pi = \pm 1$ (parity being a very well conserved
symmetry), the Hamiltonian $H$ as well as the operator ${\cal O}$ can be written
in the following block form

\begin{eqnarray}
\mbox{\begin{tabular}{ c c | c }
    \setlength{\unitlength}{1cm}
    \begin{picture}(0.2,0.0)
       \put(0.0,-0.3){\makebox(0.5,0.5){ }}
    \end{picture}
   & 
    \setlength{\unitlength}{1cm}
    \begin{picture}(3.3,0.0)
       \put(1.6,-0.3){\makebox(0.5,0.5){$+$}}
    \end{picture}
   & 
    \setlength{\unitlength}{1cm}
    \begin{picture}(2.6,0.0)
       \put(1.0,-0.3){\makebox(0.5,0.5){$-$}}
    \end{picture}
   \\
    \setlength{\unitlength}{1cm}
    \begin{picture}(0.2,2.1)
       \put(-0.1,0.1){\makebox(0.5,0.5){$+$}}
    \end{picture}
   & 
   $\begin{array}{ccc}
      \setlength{\unitlength}{1cm}
      \begin{picture}(0.3,0.3)
         \put(0.0,0.7){\makebox(0.5,0.5){ }}
      \end{picture} 
      &
      \setlength{\unitlength}{1cm}
      \begin{picture}(1.3,0.3)
         \put(0.4,0.7){\makebox(0.5,0.5){0}}
      \end{picture} 
      & 
      \setlength{\unitlength}{1cm}
      \begin{picture}(1.3,0.3)
         \put(0.4,0.7){\makebox(0.5,0.5){1}}
      \end{picture}
      \\
      \setlength{\unitlength}{1cm}
      \begin{picture}(0.3,0.5)
         \put(0.0,0.6){\makebox(0.5,0.5){0}}
      \end{picture}
      & 
      \setlength{\unitlength}{1cm}
      \begin{picture}(1.3,0.5)
          \put(0.1,0.75){\rule[-2.6mm]{1.13cm} {0.7cm}}
      \end{picture}
      & 
      \setlength{\unitlength}{1cm}
      \begin{picture}(1.3,0.5)
           \put(0.1,0.5){\framebox(1.1,0.67){ }}
           \put(0.14,0.73){.. . . }
           \put(0.34,0.78){.  .}
           \put(0.49,0.88){.   . .}
           \put(0.44,0.63){ . .}
           \put(0.74,0.58){$\ddots $}
           \put(0.79,0.88){ .}
           \put(1.04,0.83){.}
           \put(0.14,0.90){..  . . .}
           \put(0.14,0.58){. . . .}
           \put(0.14,0.98){.   . .  ..}
           \put(0.19,1.08){.  . .. .}
           \put(0.34,0.66){.}
       \end{picture}
       \\
       \setlength{\unitlength}{1cm}
       \begin{picture}(0.3,0.5)
         \put(0.0,0.2){\makebox(0.5,0.5){1}}
       \end{picture}
       & 
       \setlength{\unitlength}{1cm}
       \begin{picture}(1.3,0.5)
           \put(0.1,0.1){\framebox(1.1,0.67){ }}
           \put(0.14,0.33){.. . . }
           \put(0.34,0.38){.  .}
           \put(0.49,0.48){.   . .}
           \put(0.44,0.23){ . .}
           \put(0.74,0.18){$\ddots $}
           \put(0.79,0.48){ .}
           \put(1.04,0.43){.}
           \put(0.14,0.50){..  . . .}
           \put(0.14,0.18){. . . .}
           \put(0.14,0.58){.   . .  ..}
           \put(0.19,0.68){.  . .. .}
           \put(0.34,0.26){.}
       \end{picture}
       & 
       \setlength{\unitlength}{1cm}
       \begin{picture}(1.3,0.5)
            \put(0.1,0.35){\rule[-2.6mm]{1.13cm} {0.7cm}}
       \end{picture}
    \end{array} $
    &
    $\begin{array}{cc}
       \setlength{\unitlength}{1cm}
       \begin{picture}(1.3,0.3)
            \put(0.4,0.5){\makebox(0.5,0.5){0}}
       \end{picture} 
       &
       \setlength{\unitlength}{1cm}
       \begin{picture}(1.3,0.3)
            \put(0.4,0.5){\makebox(0.5,0.5){1}}
       \end{picture}
       \\
       \setlength{\unitlength}{1cm}
       \begin{picture}(1.3,0.5)
            \put(0.1,0.55){\framebox[1.1cm]{\rule[-1.5mm]{0cm} {0.45cm} }}
       \end{picture}
       & 
       \setlength{\unitlength}{1cm}
       \begin{picture}(1.3,0.5)
            \put(0.1,0.55){\framebox[1.1cm]{\rule[-1.5mm]{0cm} {0.45cm} }}
       \end{picture}
       \\
       \setlength{\unitlength}{1cm}
       \begin{picture}(1.3,0.5)
            \put(0.1,0.15){\framebox[1.1cm]{\rule[-1.5mm]{0cm} {0.45cm} }}
       \end{picture}
       &
       \setlength{\unitlength}{1cm}
       \begin{picture}(1.3,0.5)
            \put(0.1,0.15){\framebox[1.1cm]{\rule[-1.5mm]{0cm} {0.45cm} }}
       \end{picture}
       \\
       &
    \end{array} $ \\ 
   \hline 
  \setlength{\unitlength}{1cm}
  \begin{picture}(0.2,1.6)
       \put(-0.1,0.15){\makebox(0.5,0.5){$-$}}
    \end{picture}
  &
   $\begin{array}{ccc}
      \setlength{\unitlength}{1cm}
      \begin{picture}(0.3,0.8)
           \put(0.0,0.35){\makebox(0.5,0.5){0}}
      \end{picture}
      & 
      \setlength{\unitlength}{1cm}
      \begin{picture}(1.3,0.8)
            \put(0.1,0.55){\framebox[1.1cm]{\rule[-1.5mm]{0cm} {0.45cm} }}
       \end{picture}
       & 
       \setlength{\unitlength}{1cm}
       \begin{picture}(1.3,0.8)
            \put(0.1,0.55){\framebox[1.1cm]{\rule[-1.5mm]{0cm} {0.45cm} }}
       \end{picture}
       \\
       \setlength{\unitlength}{1cm}
       \begin{picture}(0.3,0.8)
            \put(0.0,0.35){\makebox(0.5,0.5){1}}
       \end{picture}
       &
       \setlength{\unitlength}{1cm} 
       \begin{picture}(1.3,0.8)
            \put(0.1,0.5){\framebox[1.1cm]{\rule[-1.5mm]{0cm} {0.45cm} }}
       \end{picture}
       &
       \setlength{\unitlength}{1cm}
       \begin{picture}(1.3,0.8)
            \put(0.1,0.5){\framebox[1.1cm]{\rule[-1.5mm]{0cm} {0.45cm} }}
       \end{picture}
     \end{array} $
    &
     $\begin{array}{cc} 
       \setlength{\unitlength}{1cm}
       \begin{picture}(1.3,0.8)
            \put(0.1,0.53){\rule[-2.6mm]{1.13cm} {0.7cm}}
       \end{picture}
     & 
     \setlength{\unitlength}{1cm}
     \begin{picture}(1.3,0.8)
         \put(0.1,0.27){\framebox(1.1,0.67)[c]{ }}
         \put(0.14,0.50){.. . . }
         \put(0.34,0.55){.  .}
         \put(0.49,0.65){.   . .}
         \put(0.44,0.40){ . .}
         \put(0.74,0.35){$\ddots $}
         \put(0.79,0.65){ .}
         \put(1.04,0.60){.}
         \put(0.14,0.67){..  . . .}
         \put(0.19,0.35){. . . .}
         \put(0.14,0.75){.   . .  ..}
         \put(0.19,0.85){.  . .. .}
         \put(0.34,0.43){.}
     \end{picture}
     \\ 
     \setlength{\unitlength}{1cm}
     \begin{picture}(1.3,0.8)
         \put(0.1,0.23){\framebox(1.1,0.67)[c]{ }}
         \put(0.14,0.46){.. . . }
         \put(0.34,0.51){.  .}
         \put(0.49,0.61){.   . .}
         \put(0.44,0.36){ . .}
         \put(0.74,0.31){$\ddots $}
         \put(0.79,0.61){ .}
         \put(1.04,0.56){.}
         \put(0.14,0.63){..  . . .}
         \put(0.19,0.31){. . . .}
         \put(0.14,0.71){.   . .  ..}
         \put(0.19,0.81){.  . .. .}
         \put(0.34,0.39){.}
      \end{picture}
     &    
     \setlength{\unitlength}{1cm} 
     \begin{picture}(1.3,0.8)
          \put(0.1,0.49){\rule[-2.6mm]{1.13cm} {0.7cm}}
     \end{picture}
    \end{array} $ 
\end{tabular} } .
\label{fim}
\end{eqnarray}

The four larger blocks are related to parity symmetry: The two blocks on the
diagonal correspond to transitions between positive and negative parity
configurations, respectively. The off-diagonal blocks correspond to parity
violating transitions. They are zero in the present context.

Each of the larger blocks is broken down into four smaller blocks related to
isospin symmetry. They correspond --- in a obvious way --- 
to isospin conserving and isospin violating transitions.

The ``color" of the blocks schematically shows the strength of the matrix
elements. The black blocks contain large matrix elements --- allowed by both
isospin and parity conservation. The dotted blocks contain small matrix elements
--- allowed by parity conservation but forbidden by isospin conservation. The
white blocks contain zero matrix elements --- forbidden by parity conservation.

Since in the present context parity is considered an exact symmetry, the
experimenters do not try to measure parity violating transitions. They assume
them to be zero and --- this is the important point --- they do not even list
them on their records. Hence, the experimental data set consists of the matrix
elements in the two larger diagonal blocks. They are thus equivalent to the data
that we have assumed in all foregoing considerations.

Hence, the existence of exactly conserved quantities in addition to the
partially conserved isospin quantum number is irrelevant if they are not
measured. The existence of partially conserved quantum numbers other than 
isospin would, however, alter the above results.

\section{Summary and Conclusion}
\label{VII}

In the present study, the effect of a partially broken symmetry has been
investigated for the distributions of wave function components and
transition probabilities. This has been done by means of a random matrix
model similar to the one introduced by Rosenzweig and Porter~\cite{Porter}.
However, in the present model not only the Hamiltonian but also 
the transition operator may break
the symmetry. This was dictated by the experimental material at hand and
the symmetry that we had in mind. 
The data consist of a set of reduced electromagnetic transition
probabilities in the nucleus $^{26}$Al. The symmetry is isospin
conservation. Eletromagnetic transition operators are in general not
isospin conserving.

A numerical simulation has shown that neither the components of the
eigenfunctions nor the transition matrix elements follow Gaussian
distributions. This is counterintuitive since the central limit theorem
seems to require Gaussian distributions if the wave functions are
``sufficiently complicated". Instead, the numerical simulation suggests that
the ensembles of both, the components of the eigenfunctions and the
transition matrix elements, comprise at least one class of small elements
and one class of large elements. The physical origin of these classes is
clear: The small elements are isospin violating; the large elements are
isospin allowed. At best, each of the classes has a Gaussian distribution.
The probability density of the whole ensemble is then a superposition of
Gaussians with different parameters.

This idea has been worked out in detail and results in a qualitative
analytical evaluation of the random matrix model. It yields simple formulae
for the parameters of the Gaussians. Its practical use is to circumvent easily
the demanding numerical simulations. Conceptually, it offers to
see in detail, why the premises of the central limit theorem are not given:
The elements in the different classes depend in different ways on the
symmetry breaking parameter $\alpha $ as well as the dimension $N$ of the
space. It is well known that in the limit of $N \rightarrow \infty $ an
ever smaller value of $\alpha $ suffices to make the above classes
indistinguishable and thus to produce simple Gaussian distributions of the
eigenvector components and the transition matrix elements. However, this is
what happens if $N$ is taken to infinity such that the parameter $\lambda =
\alpha /D$ goes to infinity. If $N$ is finite or if it is taken to infinity
together with $\alpha \rightarrow 0 $ such that $\lambda $ remains finite,
then the above mentioned classes conserve their identity. It is this
situation which the present paper deals with.

The present results thus support a conjecture formulated in
Ref.~\cite{26Al} on the basis of the data collected from $^{26}$Al: The
authors conjectured that the distribution of the electromagnetic transition
probabilities carries the signature of partial isospin breaking. The present
theory offers a qualitative understanding of the data. Whether it
can quantitatively reproduce the data remains to be seen. We have not
attempted to fit the theory to the data.

A caveat seems necessary: The random matrix model~(\ref{Hamiltonian}) 
applies to any partially
broken quantum number --- it is not specific for isospin. Certainly, the
available information on isospin breaking in $^{26}$Al requires that the
distribution of the transition probabilities differs from a Porter-Thomas
distribution to the extent given by the dotted curve on Fig.~\ref{fig8} ---
but the data may in principle and in addition carry the signature of other
partially conserved quantum numbers. These may not be as fundamental as
isospin but rather given by the specific form of the effective
nucleon-nucleon interaction in the sd shell.

Very recently, an analytical calculation, based on the supersymmetry
method, of the wave function statistics 
in coupled chaotic systems was presented in Ref.~\cite{TE}. These results
are valid for the case of broken time reversal invariance and therefore,
unfortunately, are not applicable to the case of conserved time reversal
invariance which is discussed in the present paper.

\acknowledgements

We thank H. A. Weidenm\"uller, A. Richter and V.K.B. Kota for fruitful
discussions.  We are indebted to G.E. Mitchell and J.F. Shriner, Jr. 
for supplying the
experimental data of $^{26}$Al and for suggestions and criticisms
during the work.  C.I.B. acknowledges the financial support granted by
the Fritz Thyssen Stiftung, Deutsche Akademische Austauschdienst
(DAAD) and Conselho Nacional de Desenvolvimento e Pesquisa (CNPq) at
different periods during this work. T.G. acknowledges support from the
Heisenberg Foundation.

\end{multicols}


\begin{references}
\bibitem{ERR}         Note that there is a lack of precision in the text of
                      \cite{nos} below Eq.(1) there: The condition of equal
                      dimensions of the GOE blocks is missing.
\end{references}
\end{document}